\documentclass[aps,prd,preprint,superscriptaddress,showpacs,longbibliography,
footnoteadded]{revtex4-1}
\usepackage[T1]{fontenc}
\usepackage{float}
\usepackage{subfigure}
\usepackage{amssymb}
\usepackage{graphicx}
\usepackage{amsmath}
\usepackage{slashed}
\usepackage{tensor}
\usepackage{hyperref}
\usepackage{epstopdf}
\usepackage{extarrows}
\usepackage{color}
\usepackage{bbm}
\usepackage{caption}%caption 宏包，加强对caption的控制
\usepackage[section]{placeins}%可以禁止图片或者表格浮动超过\section的范围
\usepackage{makecell}%表格单元格内换行
\usepackage[utf8]{inputenc}
\hypersetup{
    colorlinks=true,
    linkcolor=red,
    citecolor=blue,
}

\begin{document}

\title{Quasinormal Modes  of Analog Rotating Black Holes in 2-Dimensional Photon-Fluid Model}
\author{Hang Liu}
\email{hangliu@mail.nankai.edu.cn}
\affiliation{College of Physics and Materials Science, Tianjin Normal University, Tianjin 300387, China}

\author{Hong Guo}
\email{hong_guo@usp.br}
\affiliation{Escola de Engenharia de Lorena, Universidade de São Paulo, 12602-810, Lorena, SP, Brazil}

\author{Ru Ling}
\email{ruling@seu.edu.cn}
\affiliation{Department of Transportation, Southeast University,
Southeast University Road $\#$2, Nanjing, China, 211189}

%%%%%%%%%%%%%%%%%%%%%%%%%%%%%%%%%%%%%%%%%%%%%%%%%%%%%%%%%%%%%%%%%%%%%%%%%%%%
\begin{abstract}
It was recently found that the optical field fluctuations in self-defocusing media can be described by sound waves propagating in a two-dimensional photon-fluid. 
This photon-fluid is controlled by the driving beam and serves as the background in which the sound waves can experience an effective curved spacetime, such that it provides a new platform of studying analog black holes. 
In this paper, we are interested in investigating the quasinormal modes (QNMs) of this analog black hole in the photon-fluid model. 
Based on the master equation of motion of the optical field fluctuations, we calculate the frequencies of quasinormal modes (QNF) with three different numerical methods to make sure the QNF we get are reliable. 
Besides fundamental modes, we also try to calculate the overtones up to $n=3$ aiming to uncover more properties of QNF. 
The effects of angular velocity $\Omega_H$ of the black hole, the overtone number $n$ and the winding number $m$ on the QNF are investigated. 
Under the $m$ with opposite sign, we find that both the real and imaginary part of the QNF will show strikingly contrasting behaviors when the QNF is plotted against $\Omega_H$, and the similar contrast  effects are also found when comparing the influences from winding number and overtone number. We hope that this work may potentially contribute to the future detections of QNMs in experimental settings of photon-fluid. 
\end{abstract}

%\date{\today}

\maketitle

%%%%%%%%%%%%%%%%%%%%%%%%%%%%%%%%%%%%%%%%%%%%%%%%%%%%%%%%%%%%%%%%%%%%%%%%%%%%
\section{Introduction}

The black hole, as one of the mysterious celestial objects predicted by general relativity, remains largely enigmatic to this day. 
On the one hand, owing to breakthroughs in observational techniques, astrophysics has made significant strides in detecting black holes through both electromagnetic and gravitational wave channels.
Specifically, ground-based gravitational wave detectors like LIGO-Virgo Collaboration have observed over a hundred black hole binary merger events~\cite{LIGOScientific:2016lio,LIGOScientific:2016vlm,LIGOScientific:2018mvr}. 
Concurrently, the Event Horizon Telescope collaboration has mapped the shadow of black holes at the centers of galaxies through electromagnetic observations~\cite{EventHorizonTelescope:2019ths,EventHorizonTelescope:2019ggy,EventHorizonTelescope:2022wkp}. 
Despite these advancements, direct means to study the specific properties of black holes on a finite time scale are still lacking. 
Therefore, on the other hand, simulating black holes in laboratory environments has effectively addressed the urgent need to explore their nature~\cite{Unruh:1980cg,Unruh:1994je}. 
This endeavor is particularly remarkable as it opens possibilities for simulating complex astrophysical environments on tabletop experiments, offering further assistance in probing astrophysical and astronomical phenomena that are otherwise inaccessible. 

In 1981, Unruh first proposed the concept of an acoustic black hole in normal non-relativistic fluids and studied its process of evaporation~\cite{Unruh:1980cg}. 
Analogous to astrophysical black holes, acoustic black holes typically form when sound waves are confined within supersonic regions of a fluid. 
In such analog models of gravity, the motion equations describe the propagation of sound wave modes. 
Under certain specially engineered configurations of the fluid, spherical regions where the fluid velocity exceeds the local speed of sound are formed. 
Consequently, the boundary where the fluid velocity equals the speed of sound serves as an analog of the black hole event horizon for sound wave modes. 
This innovative setup enables the observation of relevant states of the system, facilitating the study of black hole phenomena such as event horizons, ergosphere, and Hawking radiation in laboratory settings. 

So far, a diverse array of analog black hole models has been meticulously developed, offering valuable insights into complex phenomena. At earlier times, a series of remarkable advances \cite{PhysRevA.70.063615,PhysRevA.69.033602,PhysRevLett.91.240407} in this field paved the way for studying analog gravity in ultracold quantum gases settings, followed by  recent developments \cite{Tian:2020bze,PhysRevA.106.053319,PhysRevD.105.124066,PhysRevD.107.L121502} in this direction. Besides, a latest exciting advancement in \cite{Svancara:2023yrf} which  reported observations of  rotating curved spacetime signatures from a giant quantum vortex and thus bringing us more enthusiasm  in analog gravity study.  
In 2009, the first acoustic black hole was constructed in a rubidium Bose-Einstein condensate~\cite{Lahav:2009wx}. 
More intriguingly, the remarkable experiments~\cite{MunozdeNova:2018fxv,Isoard:2019buh} reported that the thermal Hawking radiation and the corresponding temperature in an analogue black hole were observed. Additionally, recent articles on analog Hawking radiation can be found in~\cite{Anacleto:2019rfn,Balbinot:2019mei,eskin2021hawking}. 
The QNMs of 2+1-dimensional and 3+1-dimensional acoustic black holes can be found in~\cite{Visser:1997ux,Berti:2004ju,Cardoso:2004fi}, followed by some recent developments in~\cite{Daghigh:2014mwa,Torres:2020tzs,Jacquet:2021scv,Patrick:2020yyy}. 
The mechanism of black hole superradiance has also received attention~\cite{Basak:2002aw,Richartz:2009mi,Anacleto:2011tr,Patrick:2020baa}. 
Simultaneously, ongoing discussions on black hole thermodynamics, encompassing topics like black hole entropy still persist~\cite{Zhao:2012zz,Anacleto:2014apa,Anacleto:2015awa,Anacleto:2016qll}.
Moreover, the thermodynamic description of two-dimensional acoustic black holes was investigated in~\cite{Zhang:2016pqx}, while \cite{Wang:2019zqw} explores particle dynamics in acoustic black hole spacetimes. In recent years, relativistic acoustic black holes have also been constructed in Minkowski spacetime through the Abel mechanism~\cite{Ge:2010wx,Anacleto:2010cr,Anacleto:2011bv,Anacleto:2013esa} and others~\cite{Bilic:1999sq,Fagnocchi:2010sn,Visser:2010xv,Giacomelli:2017eze}.
Furthermore, building on the discussion in~\cite{Ge:2019our}, acoustic black holes in curved spacetimes are explored, including the Schwarzschild spacetime~\cite{Guo:2020blq,Qiao:2021trw,Vieira:2021ozg,Ditta:2023lny} and Reissner-Nordstr\"{o}m (RN) spacetime~\cite{Ling:2021vgk,Molla:2023hou}. For a more comprehensive understanding of analog gravity please refer to the excellent review paper~\cite{Barcelo:2005fc}.
Therefore, the significance of studying analog black holes is evident and cannot be overstated, as they have furnished invaluable tools for studying and comprehending the intricate properties of real black holes. 

Over the past decade, it was realized that rotating acoustic black holes can be generated within a self-defocusing optical cavity based on the fact that the fluid dynamics are also applicable to nonlinear optics~\cite{PhysRevA.78.063804}, thereby establishing a new platform for analog black hole research.
The basic idea of this analogy posits that when light beam propagates in the self-defocusing media, whose refractive index is intensity-dependent, such refractive index must be affected by the light beam and then the refractive index in return affects the light beam itself. 
From the perspective at microscopic level, this interaction can be perceived as a repulsive force mediated by atoms between photons, which finally leads to the formation of a “photon-fluid”. 
Leveraging the fluid-like characteristics of the optical field within the cavity, an effective curved rotating black hole spacetime can be constructed as what Unruh has done several decades ago. 
The spin of the current analog black hole can be achieved by introducing a driving light beam with a vortex profile into the cavity.
Based on the seminal work of~\cite{PhysRevA.78.063804}, acoustic superradiance of this analog black hole was subsequently investigated in~\cite{PhysRevA.80.065802}, followed by research~\cite{Ciszak:2021xlw} as a natural extension to superradiance instability (acoustic black-hole bombs).
Intriguingly, it has claimed in~\cite{Vocke2018} that this analog black hole has been experimentally constructed.
Soon after, the negative energy excitations generated by superradiance were also measured in the laboratory~\cite{Braidotti:2021nhw}.
In addition, the authors in~\cite{Braunstein:2023jpo} discussed the potential applications of the fluids system in the analog simulations of quantum gravity, which suggests a promising future of making more profound advancements in this field. 

In the present paper, we aim at investigating the QNMs of the analog rotating black hole in a 2-dimensional photon-fluid model. 
QNMs are of considerable importance in the field of black hole physics. 
Basically, QNMs can essentially be regarded as the characteristic ``sound'' of black holes~\cite{Nollert1999} as it only depends on the properties of black holes and the intrinsic properties of perturbation field, the details of how the initial perturbations are enforced are irrelevant to the QNMs. 
Accordingly, QNMs provides a rather useful and natural way to uncover the intrinsic properties of the black holes, including the analog black holes in our current consideration. 
To add a point, a related work discusses the QNMs of soliton solutions in nonlinear optics~\cite{Burgess:2023pny}.
To have a more comprehensive understanding of QNMs, one can refer to ~\cite{Konoplya:2011qq,Berti:2009kk} for a nice review.

This work is organized as follows. 
In Section~\ref{sec2}, we derive the master equation of the fluctuation field. 
In Section~\ref{sec3}, we introduce the numerical methods employed in our calculations of QNF. 
In Section~\ref{sec4}, the numerical results of QNF are demonstrated and analyzed. 
The Section~\ref{sec5}, as the last section, is devoted to conclusions and discussions.

%%%%%%%%%%%%%%%%%%%%%%%%%%%%%%%%%%%%%%%%%%%%%%%%%%%%%%%%%%%%%%%%%%%%%%%%%%%%
\section{Basic Formulas}\label{sec2}

In this section, we briefly illustrate the basic parts of this photon-fluid system.
The more details regarding connecting optical field to analog black hole spacetime are left in Appendix~\ref{Appendix}.
In order to simulate a spacetime geometry analogous to a rotating black hole metric, we choose the following background optical field profile as driving beam introduced in~\cite{PhysRevA.78.063804,PhysRevA.80.065802,Ciszak:2021xlw}
\begin{align}
E_0&=\sqrt{\rho_0}e^{i\phi_0}=\sqrt{\rho_0}e^{i(j\theta-2\pi\sqrt{r/r_0})}\\
v_r&=-\frac{c\pi}{kn_0\sqrt{r_0r}},\quad v_{\theta}=\frac{cj}{kn_0r}.
\end{align}
Considering a density perturbation $\rho_1$ of optical field in this background, it has been found that the propagation of the perturbation in this photon-fluid model is governed by Klein-Gordon equation~\cite{PhysRevA.78.063804,PhysRevA.80.065802,Ciszak:2021xlw}
\begin{equation}\label{eq=kg}
	\Box \rho_1=\frac{1}{\sqrt{-g}}\partial_{\mu}(\sqrt{-g}g^{\mu\nu}\partial_{\mu}\rho_1)=0,
\end{equation}
where $g$ is the determinant of the effective metric associated to Klein-Gordon equation. This rotating metric and some relevant black hole parameters are given by~\cite{Ciszak:2021xlw}
\begin{align}
&d s^{2} =-\left(1-\frac{r_{H}}{r}-\frac{r_{H}^{4} \Omega_{H}^{2}}{r^{2}}\right) d t^{2}+\left(1-\frac{r_{H}}{r}\right)^{-1} d r^{2}-2 r_{H}^{2} \Omega_{H} d \theta d t+r^{2} d \theta^{2},\\
&r_{H}=\frac{\xi^{2}}{r_{0}}, \Omega_{H}=\frac{j \xi}{\pi r_{H}^{2}}, \xi=\frac{\lambda}{2 \sqrt{n_{0} n_{2} \rho_{0}}},\\
&r_{\mathrm{E}}=\frac{r_{\mathrm{H}}}{2}\left(1+\sqrt{1+4 r_{\mathrm{H}}^{2} \Omega_{\mathrm{H}}^{2}}\right),
\end{align}
where $r_H$ is the event horizon, $\Omega_H$ represents the angular velocity on the event horizon, $\xi$ describes the healing length, $j$ is an integer representing the topological charge of the optical vortices, and $r_E$ stands for the radius of ergosphere.
It is worthy noting that the angular velocity $\Omega_H$ is not limited, which remarks a significant contrast to the rotating black holes (e.g, Kerr black holes) in general relativity, where $\Omega_H$ is limited in order to avoid naked singularity.  

To obtain the radial wave equation of the perturbation field, we decompose it as
\begin{equation}
	\rho_1(t,r,\theta)=R(r)e^{-i(\omega t -m \theta)},\label{eq1}
\end{equation}
where integer $m$ is called the winding number. We substitute Eq.~\eqref{eq1} into Klein-Gordon equation and a differential equation of $R(r)$ can be obtained as
\begin{align}
	&\frac{d^2R(r)}{dr^2}+P(r)\frac{dR(r)}{dr}+Q(r)R(r)=0,\label{eq2}\\
	&P_1(r)=\frac{1}{r-r_H},\\
	&Q_1(r)=\frac{r^4\omega^2-2mr^2r_H^2\omega\Omega_H-m^2(r^2-rr_H-r_H^4\Omega^2_H)}{r^2(r-r_H)^2}.
\end{align}
Now we set
\begin{equation}
R(r)=G(r)\Psi(r),
\end{equation}
and then introduce a new coordinate $r_\ast$ defined by
\begin{equation}
\frac{dr_\ast}{dr}=\Delta(r), \quad \Delta(r)=\left(1-\frac{r_H}{r}\right)^{-1}.
\end{equation}
When working in this new coordinate, Eq.~\eqref{eq2} will be transformed into 
\begin{equation}
	G(r)\Delta^2(r)\frac{d^2\Psi(r_\ast)}{dr_\ast^2}+P_2(r)\frac{d\Psi(r_\ast)}{dr_\ast}+Q_2(r)\Psi(r_\ast)=0,\label{eq3}
\end{equation}
where
\begin{align}
	P_2(r)=\frac{G(r)\Delta(r)}{r-r_H}+2\Delta(r)\frac{dG(r)}{dr}+G(r)\frac{d\Delta(r)}{dr},\\
	Q_2(r)=\frac{d^2G(r)}{dr^2}+P_1(r)\frac{dG(r)}{dr}+Q_1(r)G(r).
\end{align}
In order to obtain the Schr\"{o}dinger-like equation, the coefficient $P_2(r)$ of $d\Psi(r_\ast)/dr_\ast$ should be zero
\begin{equation}
	\frac{G(r)\Delta(r)}{r-r_H}+2\Delta(r)\frac{dG(r)}{dr}+G(r)\frac{d\Delta(r)}{dr}=0.
\end{equation}
This equation can be easily solved by following solution
\begin{equation}
G(r)=\frac{1}{\sqrt{(r-r_H)\Delta(r)}}.	
\end{equation}
Substituting $G(r)$ into Eq.~\eqref{eq3}, we finally arrive at the master equation of the perturbation field
\begin{align}
&\frac{d^2\Psi(r_\ast)}{dr^2_\ast}+U(\omega,r)\Psi(r_\ast)=0\label{mastereq},\\
&U(\omega,r)=\left(\omega-\frac{m\Omega_H}{r^2}\right)^2-\left(1-\frac{1}{r}\right)\left(\frac{m^2}{r^2}+\frac{1}{2r^3}-\frac{1}{4r^2}\left(1-\frac{1}{r}\right)\right),
\end{align}
where we have set $r_H=1$ which means that $r$ is measured in units of $r_H$, and both $\omega$ and $\Omega_H$ are measured in units of $r_H^{-1}$.
Furthermore, we may call
\begin{equation}
U(\omega,r)=\omega^2-V(\omega,r)
\end{equation}
as generalized potential and regard $V(\omega,r)$ as the effective potential, which is independent of $\omega$ in some black holes spacetime with spherical symmetry.   

%%%%%%%%%%%%%%%%%%%%%%%%%%%%%%%%%%%%%%%%%%%%%%%%%%%%%%%%%%%%%%%%%%%%%%%%%%%%
\section{The Methods}\label{sec3}

In this section, we would like to introduce three different numerical methods for calculating QNF of this analog black hole.
The rationale behind employing multiple numerical methods concurrently lies in the ability of these methods to mutually verify our numerical outcomes.
This cross-validation approach significantly enhances the reliability of our numerical results.
The black hole QNF is determined by solving the eigenvalue problem defined by Eq.~\eqref{mastereq} with the following boundary conditions for this asymptotically flat analogue black hole spacetime
\begin{equation}
\Psi \sim
\begin{cases}
   e^{-i(\omega-m \Omega_H) r_\ast}, &  r_\ast \to -\infty\quad(r\to r_H,r_H=1), \\
   e^{+i\omega r_\ast}, &  r_\ast \to +\infty\quad(r\to +\infty),
\end{cases}
\label{master_bc0}
\end{equation}
which indicates that the perturbation field is ingoing at the event horizon and outging at infinity.
The eigenvalue $\omega$ is known as the QNF, which is usually a complex value due to the dissipative nature of boundary condition Eq.~\eqref{master_bc0} that makes the differential operator of this system non-selfjoint. 

%%%%%%%%%%%%%%%%%%%%%%%%%%%%%%%%%%%%%%%%%%%%%%%%%%%%%%%%%%%%%%%%%%%%%%%%%%%%
\subsection{Asymptotic Iteration Method}

Asymptotic Iteration Method (AIM) has been widely used to calculate QNF in literatures. 
In order to employ this method, we should move back to radial coordinate $r$ under which the master equation takes the form
\begin{equation}
\frac{(r-1)^2}{r^2}\Psi''(r)+\frac{(r-1)}{r^3}\Psi'(r)+U(r) \Psi(r)=0.\label{eq4}
\end{equation}
Then for simplicity of numerical procedure, we would like to move to a new coordinate $\xi=\frac{1}{r}$. 
In this coordinate, the boundary conditions Eq.~\eqref{master_bc0} behave as
\begin{equation}
\Psi \sim
\begin{cases}
   (1-\xi)^{-i(\omega-m\Omega_H)}, &  \xi \to 1, \\
   e^{\frac{i\omega}{\xi}}\xi^{-i\omega}, &  \xi \to 0.
\end{cases}
\end{equation}
Therefore we reformulate the perturbation field $\Psi(\xi)$ as
\begin{equation}
\Psi(\xi)=	e^{\frac{i\omega}{\xi}}\xi^{-i\omega}(1-\xi)^{-i(\omega-m\Omega_H)}\chi(\xi).
\end{equation}
By the master equation, we can get
\begin{equation}
\chi''(\xi)=\lambda_{0}(\xi) \chi'(\xi)+s_{0}(\xi)\chi(\xi),
\end{equation}
where $\lambda_{0}$ and $s_{0}$ are given by
\begin{align}
&\lambda _0=-\frac{\xi  \left(-2+\xi  \left(2 i m \Omega _H-4 i \omega +3\right)\right)+2 i \omega }{(\xi -1) \xi ^2}, \\
&s_0=-\frac{4 m \left(4 (\xi +1) \omega  \Omega _H+2 i \xi  \Omega _H+m\right)+\xi  (3-16 \omega  (\omega +i))-16 \omega ^2-1}{4 (\xi -1) \xi ^2}.
\end{align}
Now the main derivations have been completed for employing AIM in numerical code, and one can refer to Ref.~\cite{Cho:2011sf} for more technical details of this method.

%%%%%%%%%%%%%%%%%%%%%%%%%%%%%%%%%%%%%%%%%%%%%%%%%%%%%%%%%%%%%%%%%%%%%%%%%%%%
\subsection{Leaver's Continued Fraction Method (CFM)}

Leaver~\cite{Leaver,PhysRevD.41.2986} found that QNF can be obtained by numerically solving a three-term recurrence relation and the so-called Leaver's method was built from then on, which is well-known for its excellent performance in QNF calculation. 
To learn more about this method please refer to \cite{Leaver,PhysRevD.41.2986,Konoplya:2011qq} where a comprehensive introduction can be found.
According to the asymptotic behavior (boundary condition Eq.~\eqref{master_bc0}) of the perturbation field $\Psi$, to which we can carry out the Frobenius series expansion around horizon as
\begin{equation}
	\Psi(r)=e^{i\omega r}r^{i\omega+i(\omega-m\Omega_H)}(r-1)^{-i(\omega-m\Omega_H)}\sum_{n=0}^{\infty}a_n \left(\frac{r-1}{r}\right)^n.\label{expansion}
\end{equation}
The next step is to derive the recurrence relation of the expansion coefficients $a_n$. 
To this end, we just need to substitute Eq.~\eqref{expansion} into Eq.~\eqref{eq4}, but before we do this, it seems necessary to work in a new coordinate $z=\frac{r-1}{r}$. 
Within this coordinate, we can get the following four-term recurrence relation
\begin{equation}
\begin{split}
&\alpha_0 a_1+\beta_0 a_0=0,\\
&\alpha_n a_{n+1}+\beta_n a_n+\gamma_n a_{n-1}=0, n\geq1,
\end{split}
\end{equation}
where
\begin{equation}
\begin{split}
& \alpha_n=4(n+1)(2 i m \Omega_H+n-2 i \omega+1)\\
& \beta_n=-2 \times\left(2 m(m+2 \Omega_H(2 i n+4 \omega+i))+(2 n-4 i \omega+1)^2\right) \\
& \gamma_n=-1+4(n-2 i \omega)(n+2 i(m \Omega_H-\omega)).
\end{split}
\end{equation}
We have shown that the expansion coefficients $a_n$ are determined by a three-term recurrence relation as above.
With this three-term recurrence relation, the QNF can be found by the condition under which the series in Eq.~\eqref{expansion} is convergent for $r\geq r_H$. 
By employing the recurrence relation, eventually the QNF can be  numerically determined by following infinite continued faction~\cite{Gautschi1967,Leaver}
\begin{equation}
\beta_0-\frac{\alpha_0 \gamma_1}{\beta_1-} \frac{\alpha_1 \gamma_2}{\beta_2-} \frac{\alpha_2 \gamma_3}{\beta_3-} \ldots \equiv \beta_0-\frac{\alpha_0 \gamma_1}{\beta_1-\frac{\alpha_1 \gamma_2}{\beta_2-\frac{\alpha_2\gamma_3}{\beta_3-\ldots}}}=0.\label{cf}
\end{equation}
Hence the QNF we are searching for are just the roots of Eq.~\eqref{cf}.

%%%%%%%%%%%%%%%%%%%%%%%%%%%%%%%%%%%%%%%%%%%%%%%%%%%%%%%%%%%%%%%%%%%%%%%%%%%%
\subsection{WKB Approximation Method}

As a semi-analytical formula, WKB approximation method is also a powerful approach for searching QNF. 
With WKB method, for a potential expressed as $U(\omega,r)=\omega^2-V(\omega,r)$, the QNF can be determined by solving following equation~\cite{Konoplya:2019hlu}
\begin{equation}
\begin{aligned}
\omega^2&=V_0(\omega)+A_2\left(\mathcal{K}^2,\omega\right)+A_4\left(\mathcal{K}^2,\omega\right)+A_6\left(\mathcal{K}^2,\omega\right)+\ldots\\
&-i \mathcal{K} \sqrt{-2 V_2(\omega)}\left(1+A_3\left(\mathcal{K}^2,\omega\right)+A_5\left(\mathcal{K}^2,\omega\right)+A_7\left(\mathcal{K}^2,\omega\right) \ldots\right),
\end{aligned}\label{eq10}
\end{equation}
where $V_0$ is the value of effective potential at its maximum $V_0=V(\omega,r_{\ast o})$ in which $r_{\ast o}$ represents the location of the peak of $V(\omega,r_\ast)$, and $V_2$ stands for the value of second order derivative of $V(\omega,r_\ast)$ respect to tortoise coordinate $r_\ast$ at the potential peak $r_{\ast o}$. 
Specifically, the terms on the right hand of Eq.~\eqref{eq10} is free of $\omega$ dependence in some cases of spherically symmetry black holes spacetime. 
We simply denote the $m$-th order derivative of $V(\omega,r_\ast)$ at $r_{\ast o}$ as $V_{m}$ given by
\begin{equation}
V_{m}=\left.\frac{d^mV(\omega,r_\ast)}{dr_\ast^m}\right|_{r_\ast=r_{\ast o}},\quad m\geq2.
\end{equation}
It is direct that $V_1=0$, and $A_{k}(\mathcal{K}^2,\omega)$ are polynomials of $V_2,V_3,\ldots V_{2k}$, and each $A_{k}(\mathcal{K}^2,\omega)$ should be considered as the $k$-th order corrections to the eikonal formula
\begin{equation}
\mathcal{K}=i\frac{\omega^2-V_0}{\sqrt{-2V_2}},	
\end{equation}
which provides an unique solution for $\mathcal{K}$ with a given $\omega$. 
With the boundary conditions of QNMs, it is found that $\mathcal{K}$ has to be constrained as 
\begin{equation}
\mathcal{K}=n+\frac{1}{2},\quad n\in\mathbb{N},	
\end{equation}
in which $n$ is the overtone number.

With these formulas in hand, we are going to calculate QNF with $6$-th order WKB approximation method. 
Here we list the second and third order corrections as follows
\begin{equation}
	A_2(\mathcal{K}^2,\omega)=\frac{-60\left(n+\frac{1}{2}\right)^2 V_3^2+36\left(n+\frac{1}{2}\right)^2 V_2 V_4-7 V_3^2+9 V_2 V_4}{288 V_2^2},
\end{equation}
\begin{equation}
\begin{aligned}
A_3(\mathcal{K}^2,\omega)&=\frac{1}{13824 V_2^5}\Bigg[-940\left(n+\frac{1}{2}\right)^2 V_3^4+1800\left(n+\frac{1}{2}\right)^2 V_2 V_4 V_3^2-672\left(n+\frac{1}{2}\right)^2 V_2^2 V_5 V_3\\
& -204\left(n+\frac{1}{2}\right)^2 V_2^2 V_4^2+96\left(n+\frac{1}{2}\right)^2 V_2^3 V_6-385 V_3^4+918 V_2 V_4 V_3^2-456 V_2^2 V_5 V_3 \\
&  -201 V_2^2 V_4^2+120 V_2^3 V_6\Bigg].
\end{aligned}
\end{equation}
For the remaining higher order corrections $A_4,A_5$ and $A_6$, which are too complex to be demonstrated explicitly, one can refer to~\cite{Konoplya:2019hlu} and references therein for explicit expressions.

%%%%%%%%%%%%%%%%%%%%%%%%%%%%%%%%%%%%%%%%%%%%%%%%%%%%%%%%%%%%%%%%%%%%%%%%%%%%
\section{Numerical Results of QNF}\label{sec4}

In this section we demonstrate and discuss the numerical results of QNF. 
As at the beginning of this section, we should point out that for the winding $m=0$ and $m<0$ with small magnitude, the capability of all numerical methods to obtain reliable results is highly limited. 
For this reason, we only focus on the QNF with $m=-10,-5,1,5$ and $m=10$ in our following discussions in order to have accurate descriptions for the characteristics of QNF. 

In Table.~\ref{tab1}, we list the QNF with $m=1$ for different angular velocity $\Omega_H$. 
For each angular velocity, we calculated the first four overtones up to $n=3$, meanwhile the numerical results obtained by AIM and CFM are presented together in order to verify the accuracy of our QNF calculations. 
It is satisfying to observe that the data in the table show a high degree of concordance between the numerical results derived from these two methods. 
From this table, one can observe that for a given overtone number, the real part $\omega_R$ of QNF grows rapidly with the increase of angular velocity, manifesting a sensitive response to the change of $\Omega_H$. 
On the other hand, since $\omega_R$ stands for the oscillation frequency of QNMs, it suggests that the black hols with higher spin will support QNMs which oscillate more rapidly in this analogue black hole background. 
Conversely, for a fixed $\Omega_H$, the observation is that higher overtone corresponds to a lower $\omega_R$.
This indicates that changes in the overtone number result in a more moderate variation of $\omega_R$ compared to the variations induced by changes in $\Omega_H$.
For the imaginary part $\omega_I$ of QNF, which characterizes the damping rate of QNMs, behaves drastically different from the $\omega_R$. 
On the contrary to the real part, $\omega_I$ is sensitive to the change of overtone number while it is weakly dependent on $\Omega_H$. 
By increasing the angular velocity, the magnitude of the negative $\omega_I$ will become bigger, which means that the QNMs will fade away more rapidly in a higher spin analogue black hole background. 
For higher overtones, $\omega_I$ naturally possess a larger magnitude, indicating a much shorter life for QNMs. 
This underscores why the $n=0$ mode is often referred to as the fundamental mode, simply because it exhibits a much longer duration compared to the higher overtones.
It is also interesting to note that for larger $\Omega_H$, the difference in $\omega_R$ between the overtones becomes less pronounced, suggesting an asymptotically linear relationship with $\Omega_H$.
While for $\omega_I$, sufficiently high angular velocity seems to render the disparities between adjacent $\omega_I$ approach to a constant around $\thicksim 0.5$, and for each overtone it appears to approach to a constant value.

%%%%%%%%%%%%%%%%%%%%%%%%%%%%%%%
\begin{table}[!htbp]
\centering
\caption*{$m=1$}
\resizebox{\textwidth}{!}
{
        \begin{tabular}{ccccccc}
    \hline\hline
    $n$ &Method&  $\Omega_H=0$  &$\Omega_H=0.5$          & $\Omega_H=1$          & $\Omega_H=5$             & $\Omega_H=10$       \\
    \hline
    $0$ &AIM&     $0.365926-0.193965i$           & $ 0.676790-0.222215i $ & $ 1.10290-0.237129i $ & $ 5.02182-0.249306i $    & $ 10.0109-0.249825i $\\
        \cline{2-7}
        &CFM&     $0.365932-0.193959i$           & $ 0.676776-0.222225i $ & $ 1.10290-0.237130i $ & $ 5.02182-0.249306i $    & $ 10.0109-0.249825i $\\
        \hline
    $1$ &AIM&     $0.295643-0.633857i$           & $ 0.646675-0.684320i $ & $ 1.09101-0.716744i $ & $ 5.02161-0.747943i $    & $ 10.0109-0.749476i $\\
        \cline{2-7}
        &CFM&     $0.295391-0.636114i$           & $ 0.647186-0.684593i $ & $ 1.09107-0.716704i $ & $ 5.02161-0.747943i $    & $ 10.0109-0.749476i $\\
        \hline
    $2$ &AIM&     $0.240786-1.12826i$            & $ 0.614894-1.17116i $  & $ 1.07548-1.20601i $  & $ 5.02123-1.24665i $     & $ 10.0109-1.24913i $\\
        \cline{2-7}
        &CFM&     $0.246357-1.11459i$            & $ 0.617152-1.16650i $  & $ 1.07511-1.20540i $  & $ 5.02123-1.24665i $     & $ 10.0109-1.24913i $\\
        \hline
    $3$ &AIM&     $0.206897-1.63312i$            & $ 0.591555-1.66908i $  & $ 1.06181-1.70195i $  & $ 5.02068-1.74547i $     & $ 10.0108-1.74880i $\\
        \cline{2-7} 
        &CFM&     $0.21751-1.66256i$             & $ 0.596942-1.64467i $  & $ 1.06083-1.69653i $  & $ 5.02068-1.74547i $     & $ 10.0108-1.74880i $\\
         \hline\hline
\end{tabular}
}
\caption{The first four overtones $n=0,1,2,3$ of QNF at $m=1$ for different $\Omega_H$. The numerical results obtained by AIM and CFM are presented together for comparison to verify the validity of the calculation methods.}\label{tab1}
\end{table}
%%%%%%%%%%%%%%%%%%%%%%%%%%%%%%%

\begin{figure}[thbp]
\centering
\includegraphics[height=2.4in,width=3.2in]{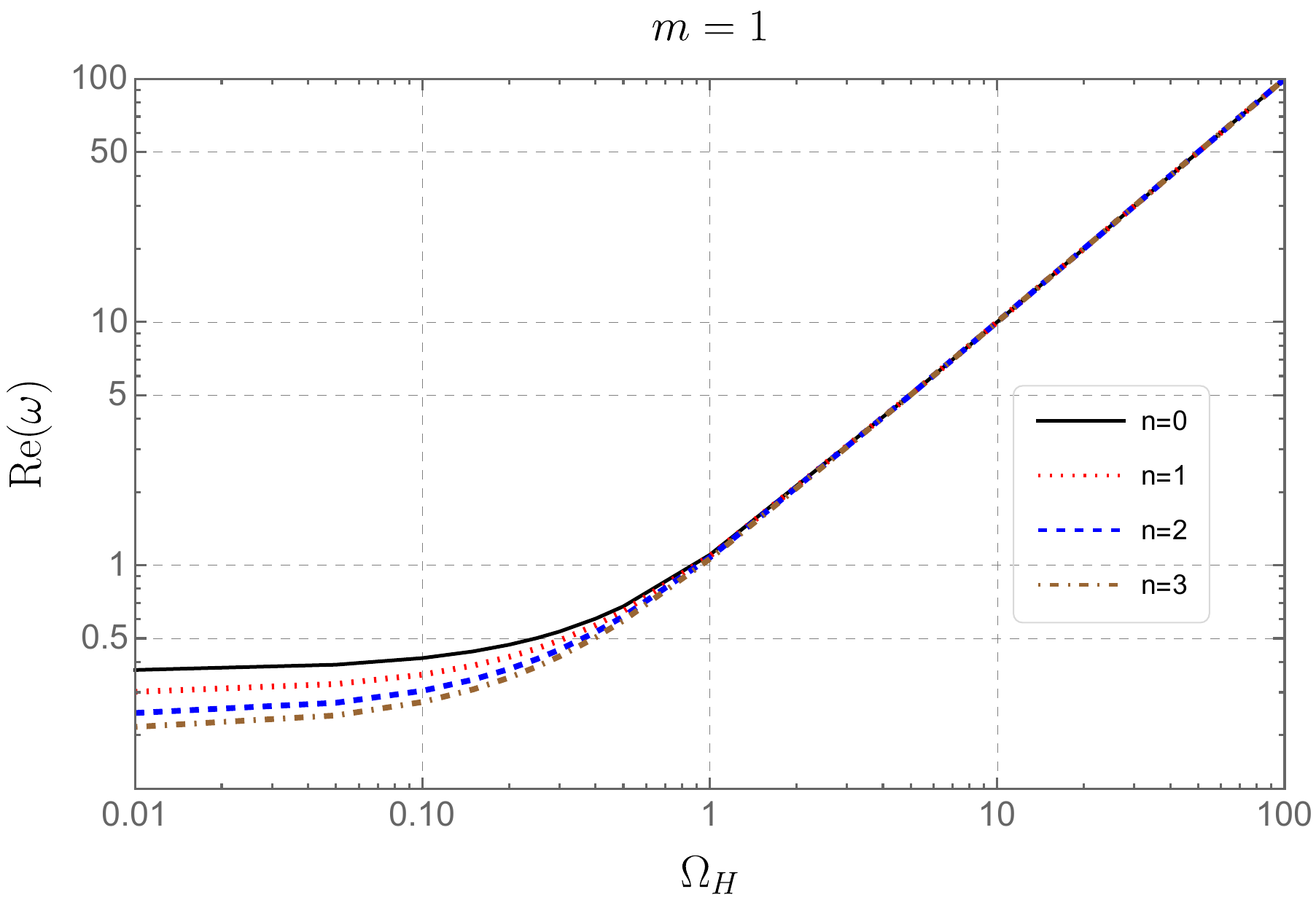}
\includegraphics[height=2.4in,width=3.2in]{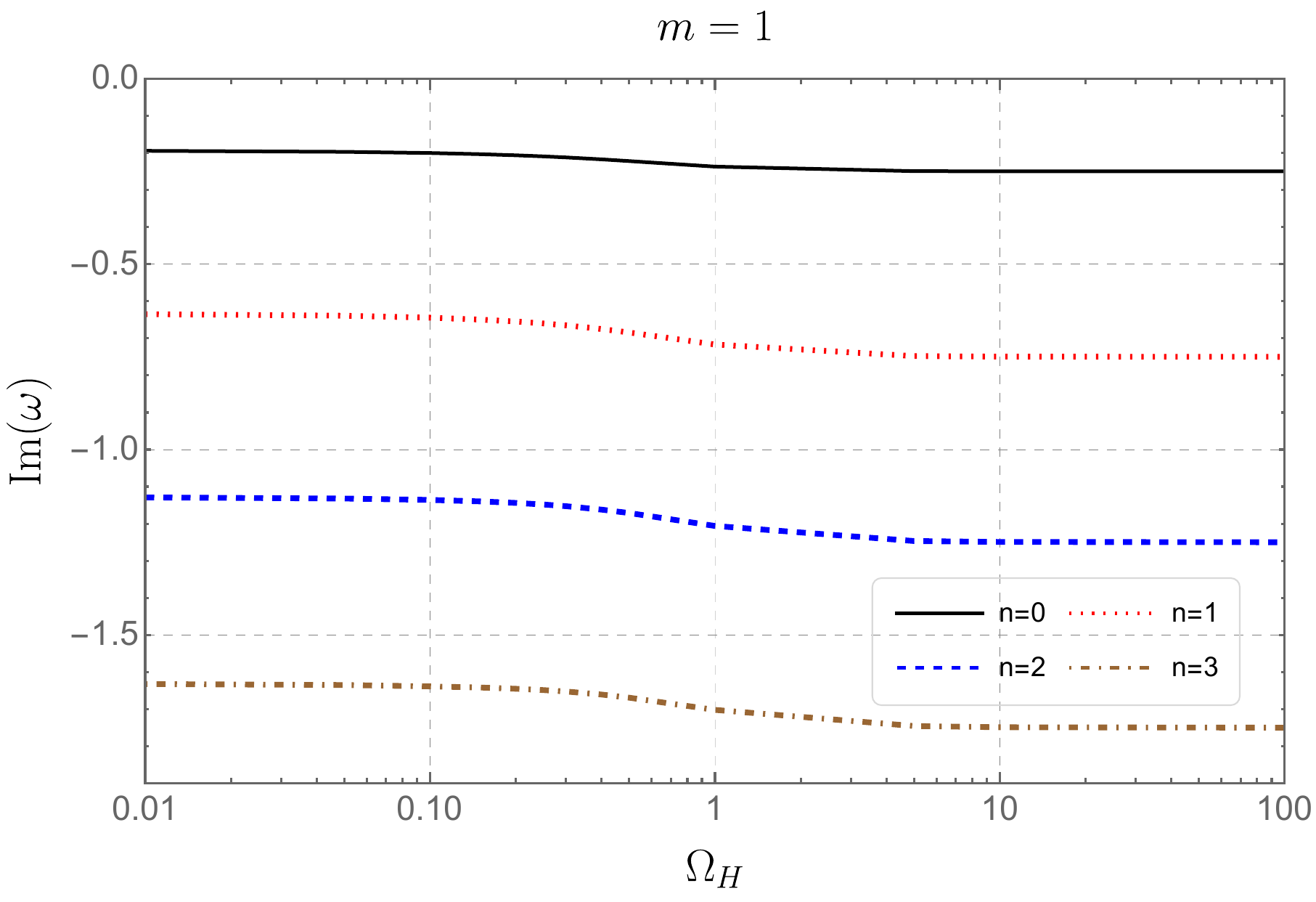}
\caption{The dependence of real part $\omega_R$ (left plot) and imaginary part $\omega_I$ (right plot) of QNF on $\Omega_H$ for $m=1$. In each plot, we demonstrate four overtones from $n=0$ to $n=3$.\label{fig1}}
\end{figure}

To provide a direct demonstration of the properties of the QNF, we plot the QNF data from Table.~\ref{tab1} as a function of $\Omega_H$ in Fig.~\ref{fig1}, where the left and right plots depict the behavior of $\omega_R$ and $\omega_I$ under the change of $\Omega_R$, respectively. 
In the plots, the range of angular velocity has been extended to $\Omega_H=100$ with the purpose to illustrate the characteristics of the QNF as clear as possible. 
All the features of the QNF data in Table.~\ref{tab1} are directly presented in the figure, which additionally shows that the imaginary part of fundamental mode is the most insensitive to the angular velocity variation.

%%%%%%%%%%%%%%%%%%%%%%%%%%%%%%%
%%%%%%%%%%%%%%%%%%%%%%%%%%%%%%%

\begin{figure}[thbp]
\centering
\includegraphics[height=2.4in,width=3.2in]{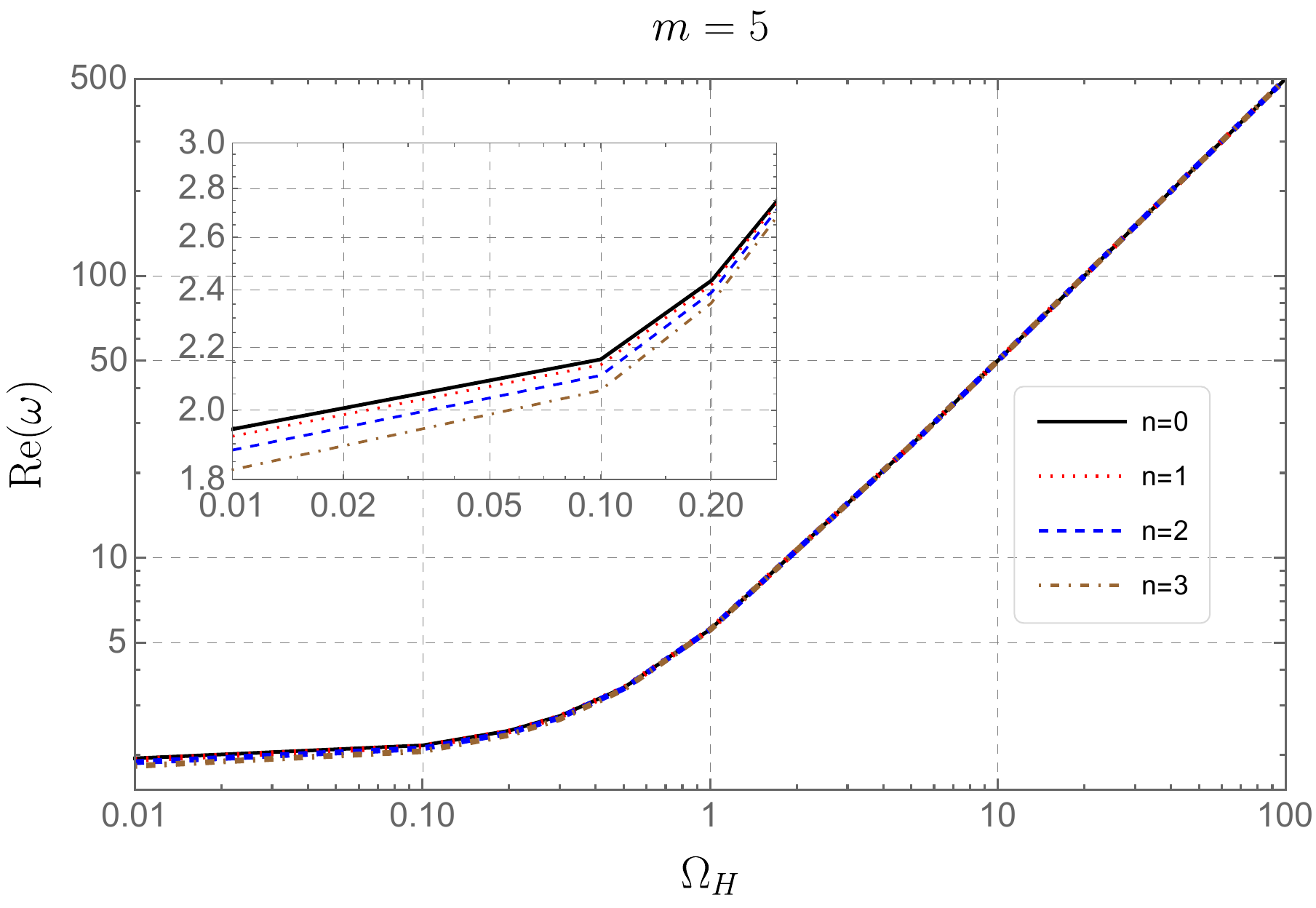}
\includegraphics[height=2.4in,width=3.2in]{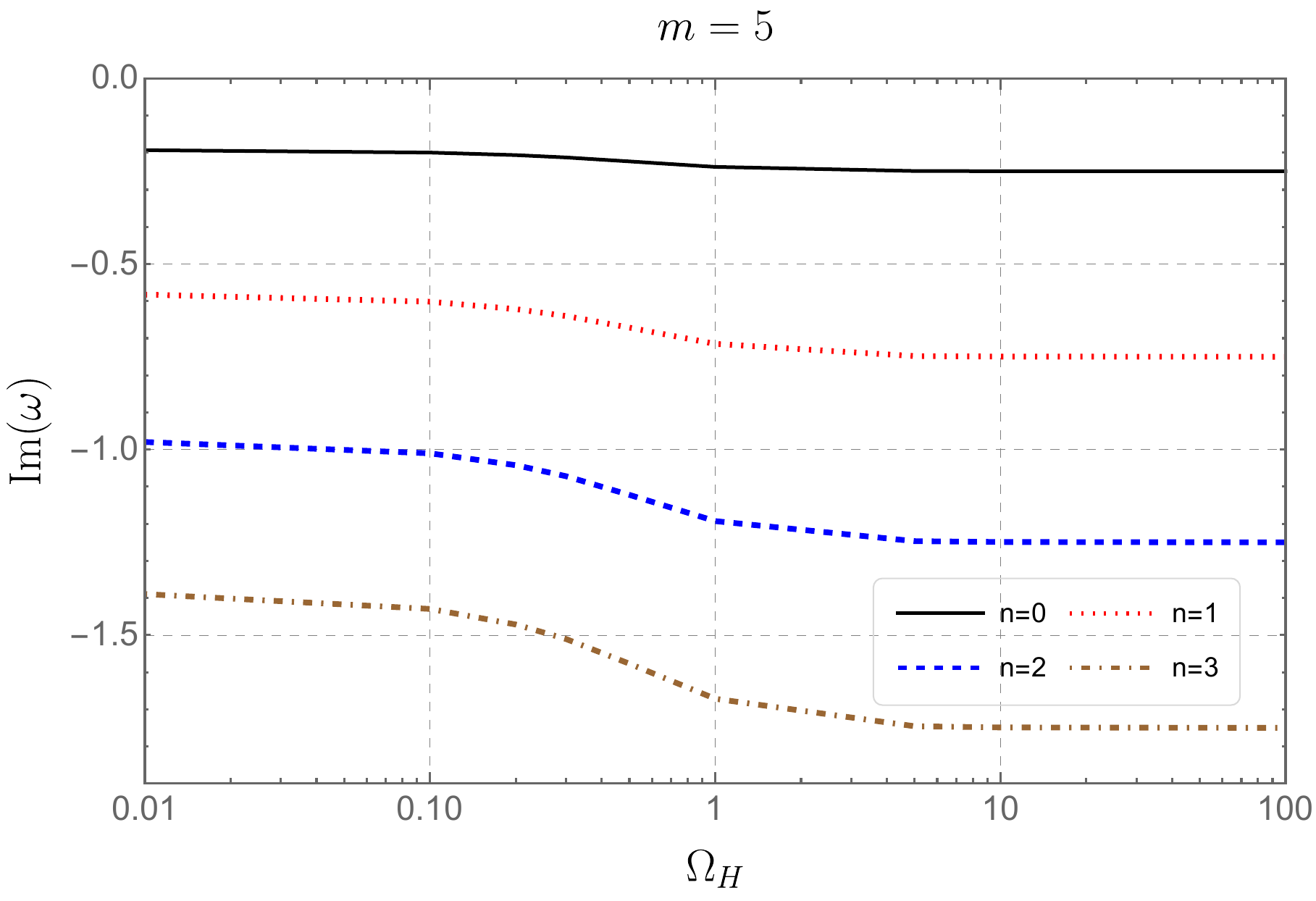}
\caption{The dependence of real part $\omega_R$ (left plot) and imaginary part $\omega_I$ (right plot) of QNF on $\Omega_H$ for $m=5$. In each plot, we demonstrate four overtones from $n=0$ to $n=3$.\label{fig2}}
\end{figure}

\begin{figure}[thbp]
\centering
\includegraphics[height=2.4in,width=3.2in]{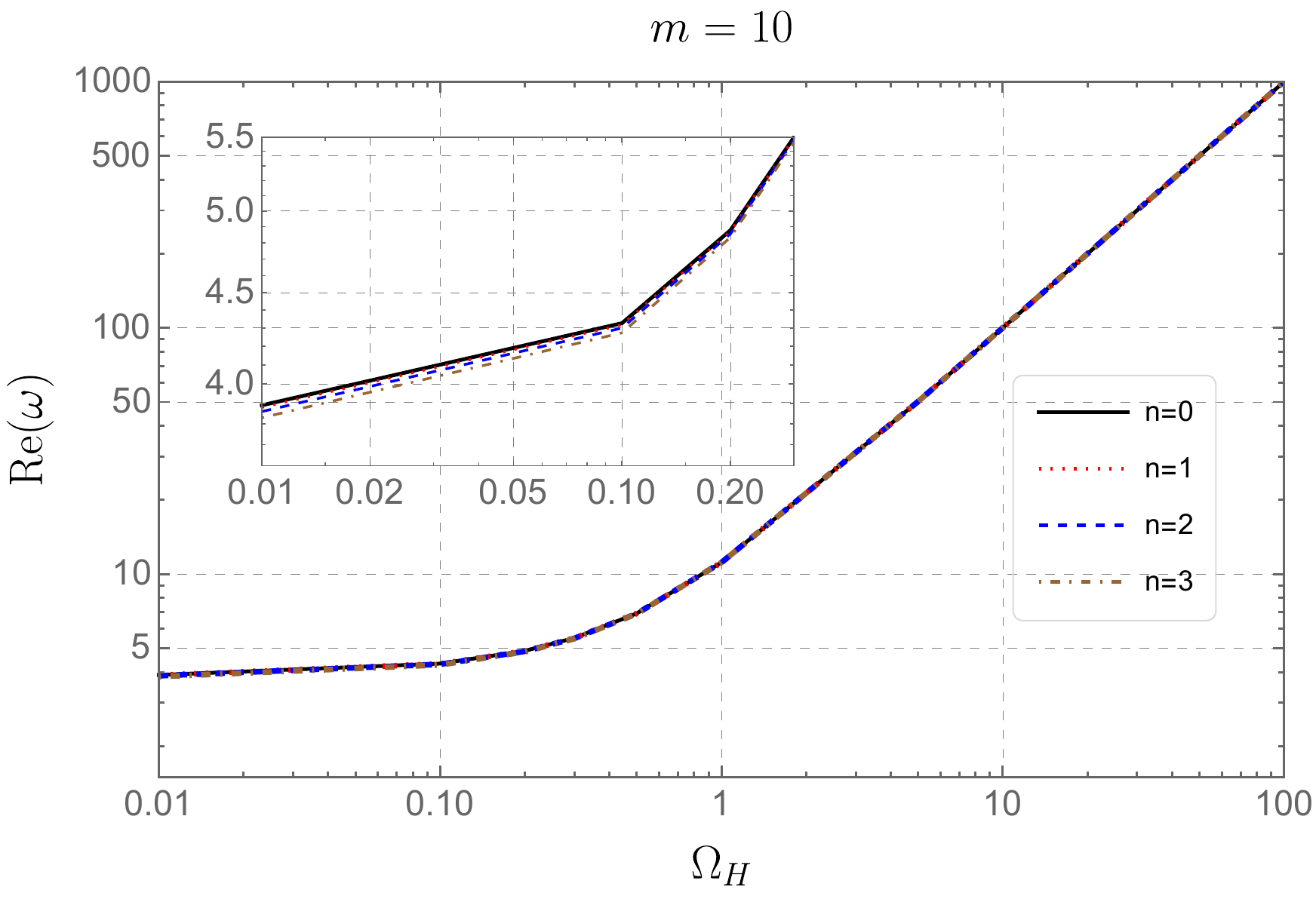}
\includegraphics[height=2.4in,width=3.2in]{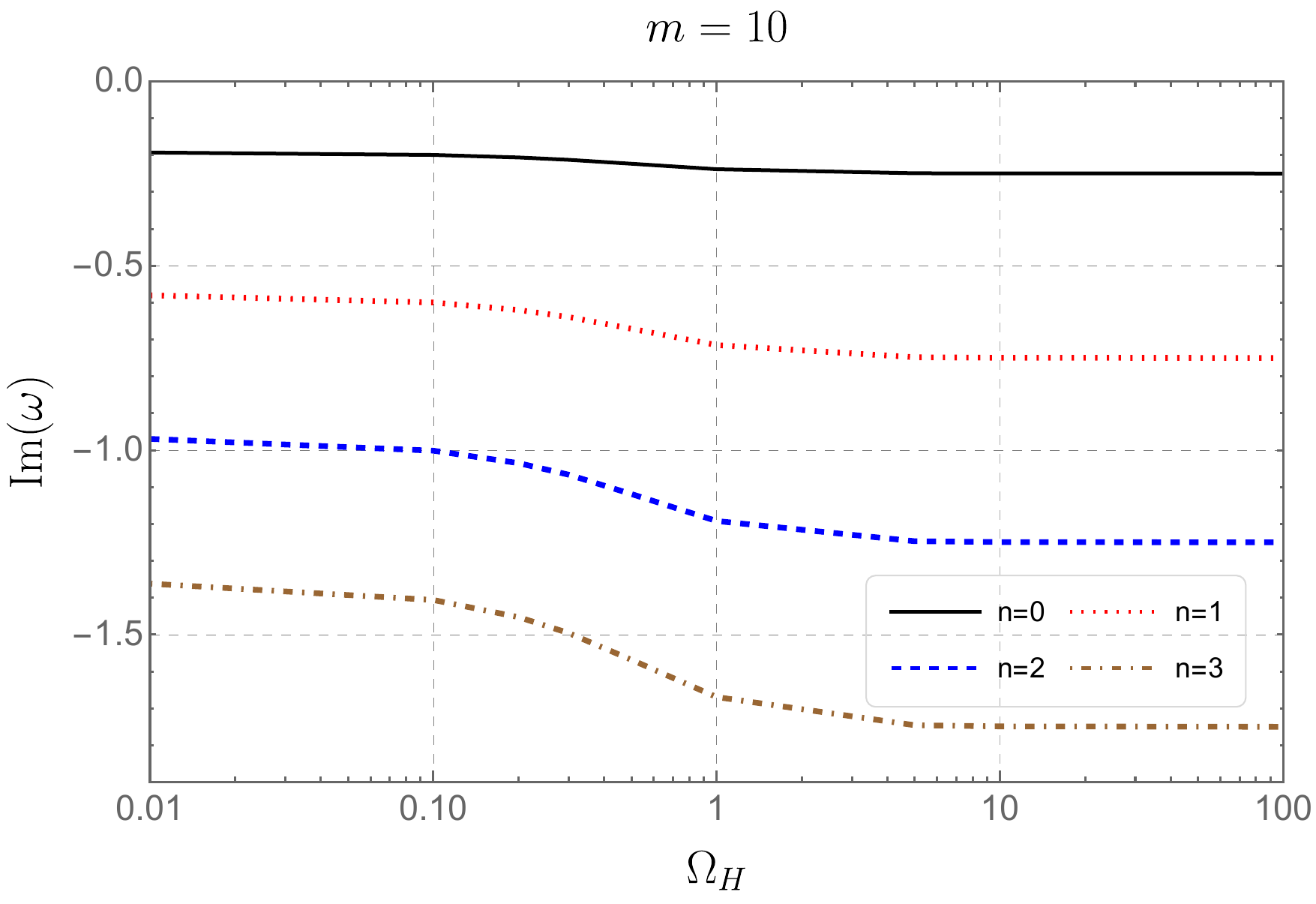}
\caption{The dependence of real part $\omega_R$ (left plot) and imaginary part $\omega_I$ (right plot) of QNF on $\Omega_H$ for $m=10$. In each plot, we demonstrate four overtones from $n=0$ to $n=3$.\label{fig3}}
\end{figure}

We demonstrate the numerical results of QNF for $m=5$ in Table.~\ref{tab2} and Fig.~\ref{fig2}, and for $m=10$ in Table.~\ref{tab3} and Fig.~\ref{fig3}. 
The behaviors of the QNF in this two cases are qualitatively similar to that of case $m=1$, except that for these higher values of $m$, $\omega_R$ is much more sensitive to the changes in $\Omega_H$, which also leads to a slightly larger amplitude of variation in $\omega_I$. 
While for the fundamental mode, the imaginary part still remains as the least susceptible to the changes in angular velocity.
Interestingly, from the tables, we can observe that the differences in $\Omega_I$ between adjacent overtones get close to $\Delta \omega_I\approx 0.5$ as what we have found in case of $m=1$. 
By summarizing the QNF in the three specific cases of positive $m$, we may conclude that the qualitative features of QNF discovered presently are consistent across all positive values of $m$.  

%%%%%%%%%%%%%%%%%%%%%%%%%%%%%%%
\begin{table}[!htbp]
\centering
\caption*{$m=5$}
\resizebox{\textwidth}{!}
{
        \begin{tabular}{ccccccc}
    \hline\hline
    $n$ &Method&  $\Omega_H=0$         &$\Omega_H=0.5$       & $\Omega_H=1$        & $\Omega_H=5$           & $\Omega_H=10$       \\
    \hline
    $0$ &AIM&     $1.92059-0.192507i$  & $3.45759-0.223514i$ & $5.57214-0.238330i$ & $25.1239-0.249380i$    & $50.0621-0.249844i$\\
        \cline{2-7}
        &CFM&     $1.92059-0.192507i$  & $3.45759-0.223514i$ & $5.57214-0.238330i$ & $25.1239-0.249380i$    & $50.0621-0.249844i$\\
        \hline
    $1$ &AIM&     $1.89952-0.580296i$  & $3.44961-0.671489i$ & $5.56945-0.715239i$ & $25.1239-0.748142i$    & $50.0621-0.749531i$\\
        \cline{2-7}
        &CFM&     $1.89952-0.580296i$  & $3.44961-0.671489i$ & $5.56945-0.715239i$ & $25.1239-0.748142i$    & $50.0621-0.749531i$\\
        \hline
     $2$ &AIM&    $1.85899-0.976231i$  & $3.43424-1.12220i$  & $5.56421-1.19288i$  & $25.1238-1.24691i$     & $50.0621-1.24922i$\\
        \cline{2-7}
        &CFM&     $1.85899-0.976232i$  & $3.43424-1.12219i$  & $5.56421-1.19288i$  & $25.1238-1.24691i$     & $50.0621-1.24922i$\\
        \hline
    $3$ &AIM&     $1.80226-1.38504i$   & $3.41251-1.57715i$  & $5.55664-1.67167i$  & $25.1237-1.74568i$     & $50.0621-1.74891i$\\
        \cline{2-7}
        &CFM&     $1.80226-1.38505i$   & $3.41251-1.57716i$  & $5.55664-1.67167i$  & $25.1237-1.74568i$     & $50.0621-1.74891i$\\
         \hline\hline
\end{tabular}
}
\caption{The first four overtones $n=0,1,2,3$ of QNF at $m=5$ for different $\Omega_H$. The numerical results obtained by AIM and CFM.}\label{tab2}
\end{table}
%%%%%%%%%%%%%%%%%%%%%%%%%%%%%%%

%%%%%%%%%%%%%%%%%%%%%%%%%%%%%%%
%%%%%%%%%%%%%%%%%%%%%%%%%%%%%%%

%%%%%%%%%%%%%%%%%%%%%%%%%%%%%%%
\begin{table}[!htbp]
\centering
\caption*{$m=10$}
\resizebox{\textwidth}{!}
{
        \begin{tabular}{ccccccc}
    \hline\hline
    $n$ &Method&  $\Omega_H=0$  &$\Omega_H=0.5$          & $\Omega_H=1$          & $\Omega_H=5$             & $\Omega_H=10$       \\
    \hline
    $0$ &AIM&     $3.84704-0.192464i$  & $6.92012-0.223568i$ & $11.1480-0.238370i$ & $50.2488-0.249383i$    & $100.125-0.249844i$\\
        \cline{2-7}
        &CFM&     $3.84704-0.192464i$  & $6.92012-0.223568i$ & $11.1480-0.238370i$ & $50.2488-0.249383i$    & $100.125-0.249844i$\\
        \hline
    $1$ &AIM&     $3.83639-0.578090i$  & $6.91609-0.670944i$ & $11.1467-0.715172i$ & $50.2487-0.748148i$    & $100.125-0.749532i$\\
        \cline{2-7}
        &CFM&     $3.83639-0.578090i$  & $6.91609-0.670944i$ & $11.1467-0.715172i$ & $50.2487-0.748148i$    & $100.125-0.749532i$\\
        \hline
    $2$ &AIM&     $3.81528-0.965797i$  & $6.90811-1.11903i$ & $11.1440-1.19216i$ & $50.2487-1.24691i$       & $100.125-1.24922i$\\
        \cline{2-7}
        &CFM&     $3.81528-0.965797i$  & $6.90811-1.11903i$ & $11.1440-1.19216i$ & $50.2487-1.24691i$       & $100.125-1.24922i$\\
        \hline
    $3$ &AIM&     $3.78412-1.35694i$  & $6.89632-1.56828i$ & $11.1400-1.66946i$ & $50.2486-1.74568i$        & $100.125-1.74891i$\\
        \cline{2-7}
        &CFM&     $3.78412-1.35694i$  & $6.89632-1.56828i$ & $11.1400-1.66946i$ & $50.2486-1.74568i$        & $100.125-1.74891i$\\
         \hline\hline
\end{tabular}
}
\caption{The first four overtones $n=0,1,2,3$ of QNF at $m=10$ for different $\Omega_H$. The numerical results obtained by AIM and CFM.}\label{tab3}
\end{table}
%%%%%%%%%%%%%%%%%%%%%%%%%%%%%%%

%%%%%%%%%%%%%%%%%%%%%%%%%%%%%%%
%%%%%%%%%%%%%%%%%%%%%%%%%%%%%%%

\begin{figure}[thbp]
\centering
\includegraphics[height=2.4in,width=3.15in]{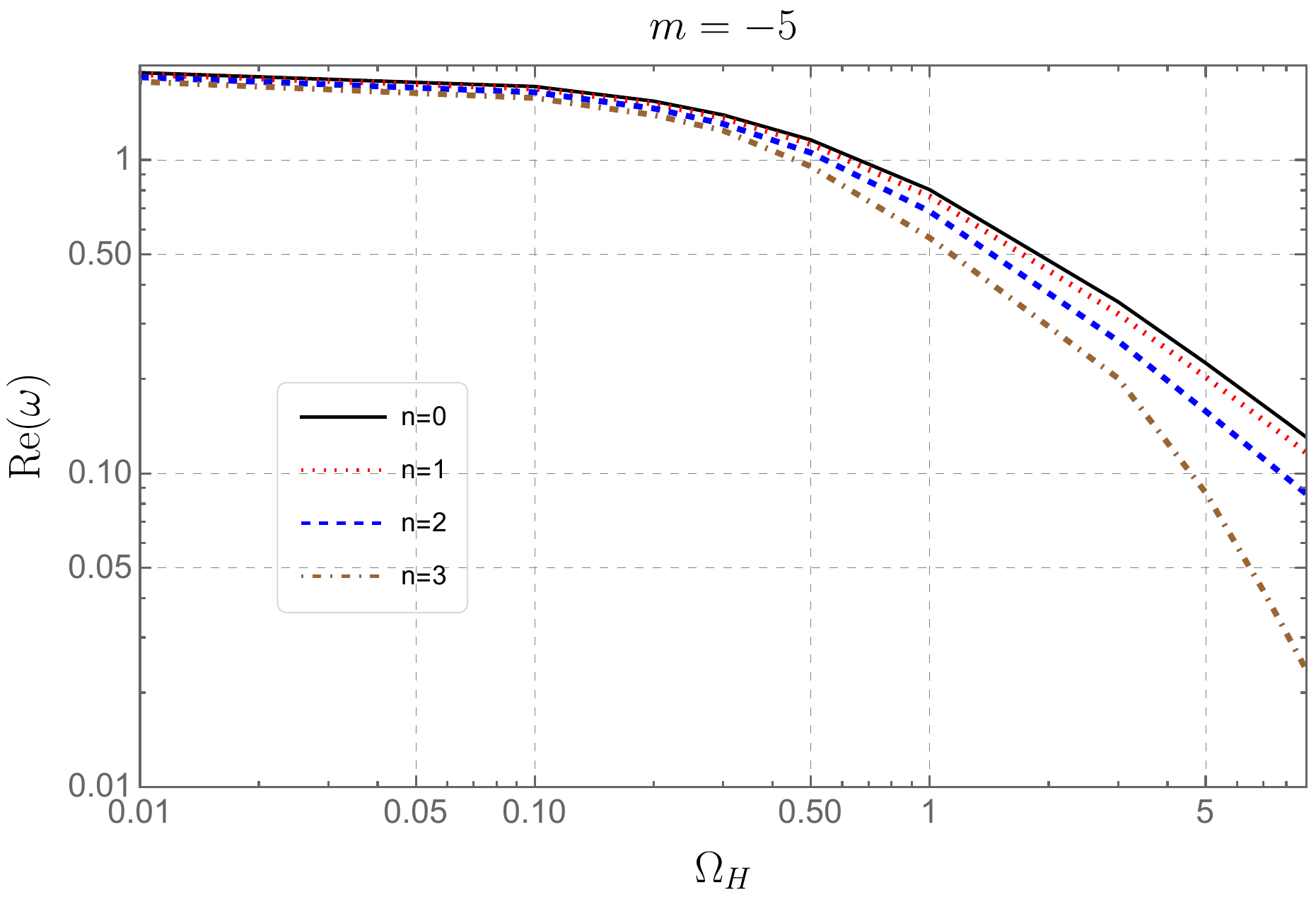}
\includegraphics[height=2.4in,width=3.15in]{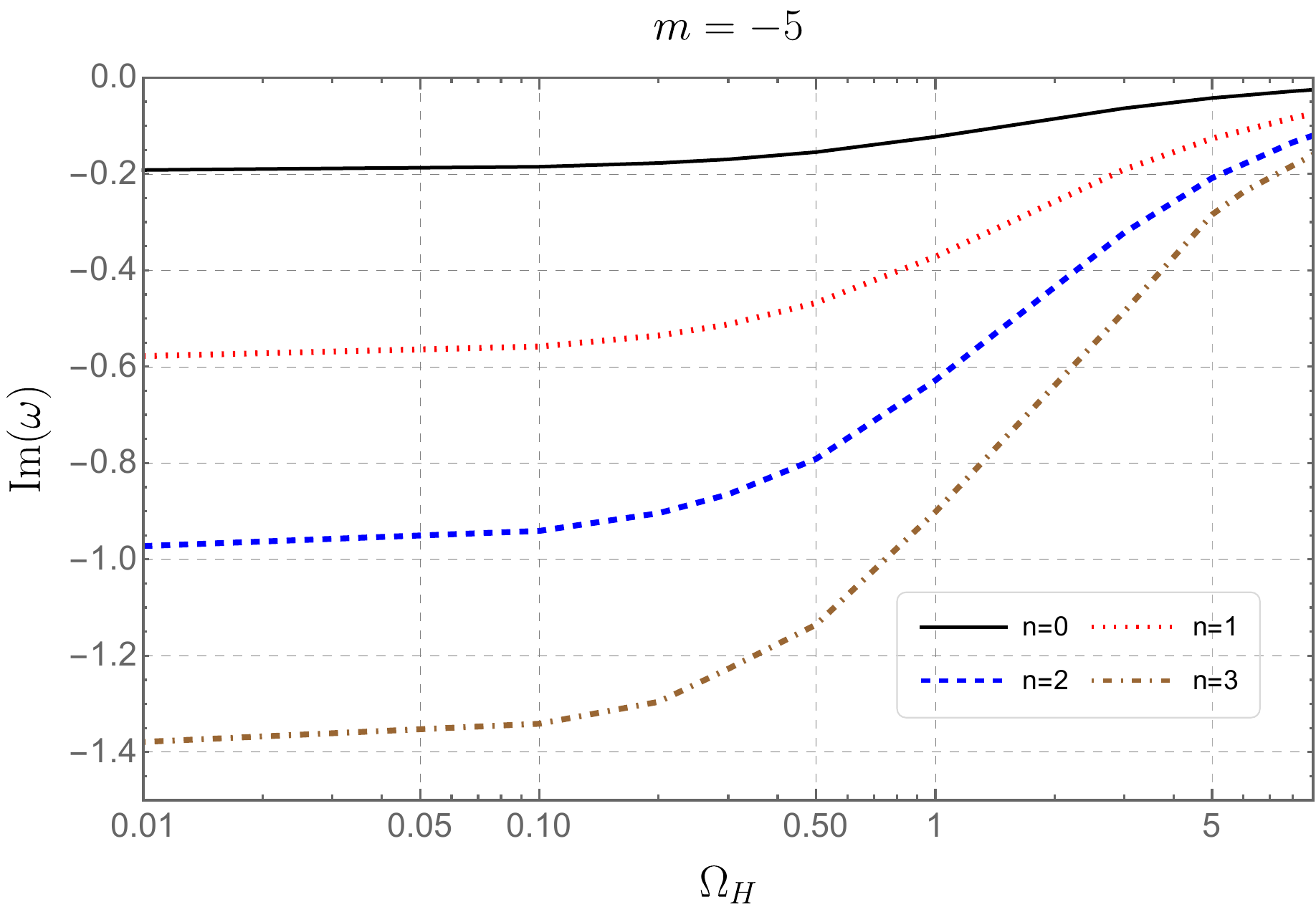}

\includegraphics[height=2.4in,width=3.2in]{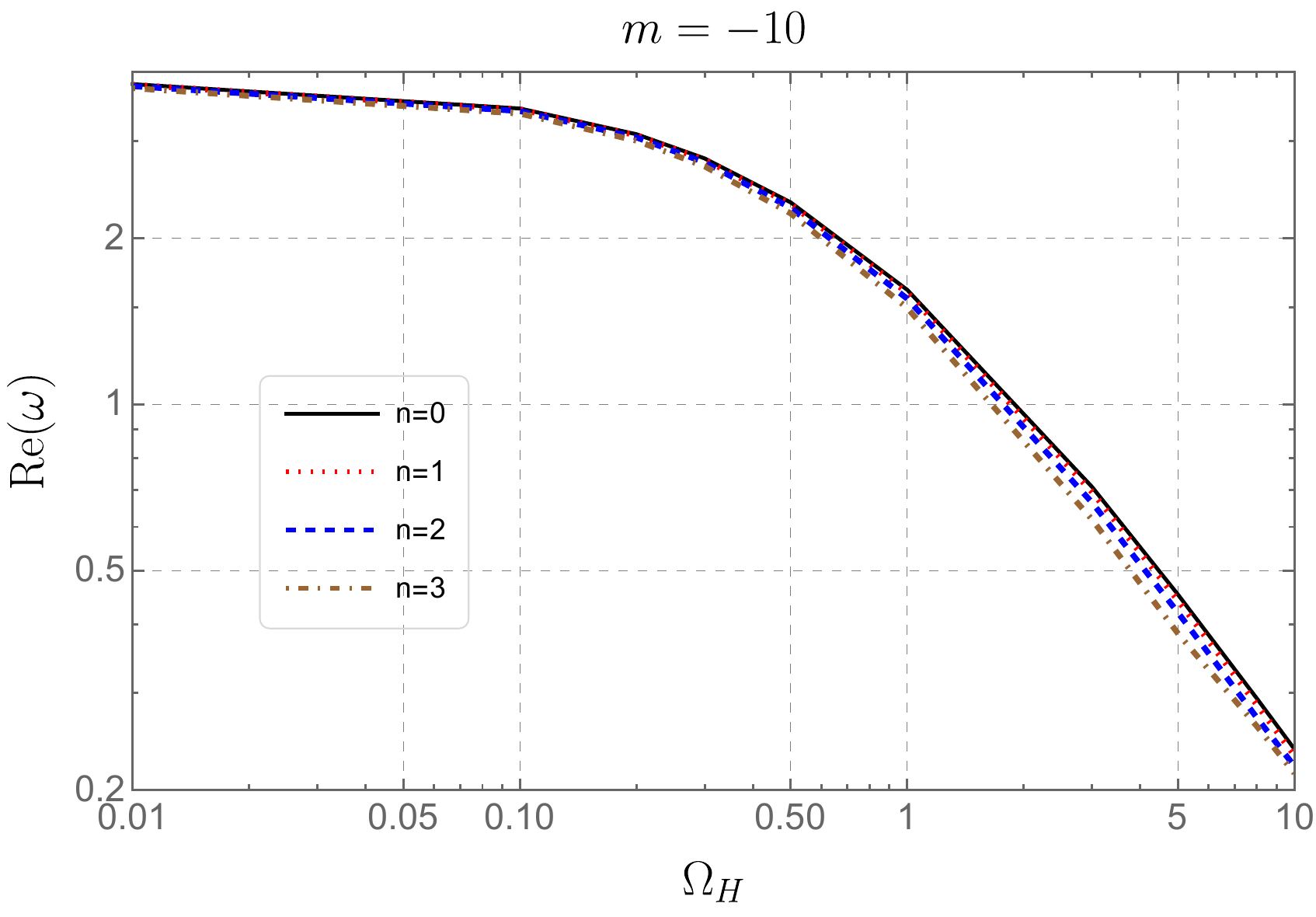}
\includegraphics[height=2.4in,width=3.2in]{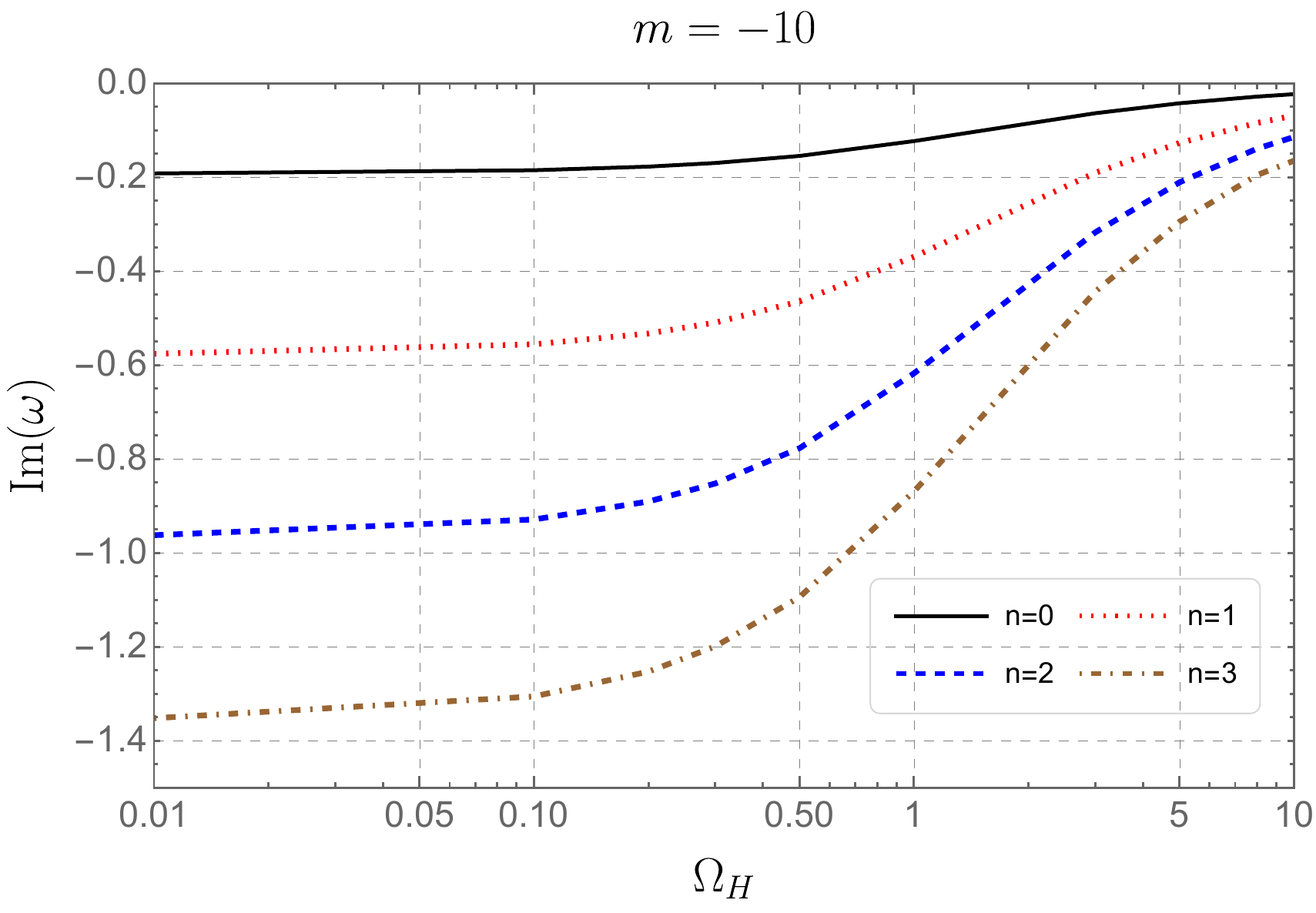}
\caption{The dependence of real part $\omega_R$ (left plots) and imaginary part $\omega_I$ (right plots) of QNF on $\Omega_H$ for $m=-5$ (upper panel) and $m=-10$ (bottom panel). In each plot, we demonstrate four overtones from $n=0$ to $n=3$. Note that for $m=-5$, all the numerical methods failed to output reliable  results for overtone $n=3$ at $\Omega_H=10$, we narrow down the range of $\Omega_H$ to $(0,9)$ in the plots to guarantee accuracy.  \label{fig4}}
\end{figure}

After discussing the properties of QNF with positive $m$, now we turn to the case of negative $m$. 
However, in this scenario, our numerical methods limit us to sufficiently large $|m|$ and small $\Omega_H$ (we take $0\leqslant\Omega_H\leqslant10$), otherwise we can not obtain reliable numerical results for overtones, although the QNF of fundamental modes ($n=0$) are still available. 
Therefore, we take $m=-5$ and $m=-10$ as  examples to illustrate the characteristics of QNF for a negative winding number. In the case of $m=-5$, the AIM does  not work well so here we adopt six order WKB approximation method, which turns out to yield credible values of QNF, in association with CFM to calculate QNF already presented in Table.~\ref{tab5}.
When $m=-10$, the AIM is now capable of calculating QNF accurately, so this time we use AIM together with six order WKB approximation method to obtain QNF which are demonstrated in Table.~\ref{tab4}. The  QNF for both $m=-5$ and $m=-10$ are also displayed by pictures  in  Fig.~\ref{fig4} which shows a qualitatively similar behavior for these two negative winding number. 
Interestingly, a remarkably different QNF behaviors from what we have discussed for positive $m$ cases can be directly observed in the plots, although some common characteristics remain untouched, such as for a fixed $\Omega_H$, $\omega_R$ is only mildly decreased by increasing overtone number $n$, while $\omega_I$ suffers a considerable change.
On the contrary to the QNF with $m>0$, where $\omega_R$ grows with angular velocity, in present case, $\omega_R$ rapidly drops off when increasing $\Omega_H$. 
For $\omega_I$, it is now sensitive to the changes in $\Omega_H$, rather than being insensitive as in positive $m$ case. 
With the increase of $\Omega_H$, the $\omega_I$ of all overtones grows and appears to converge to a small constant instead of approaching $|\Delta\omega_I|\approx 0.5$ between two adjacent overtones in the $m>0$ case. 
This result may suggest the existence of arbitrarily long-lived QNMs which is also called quasi-resonances~\cite{Churilova:2019qph,Konoplya:2004wg,Ohashi:2004wr} when angular velocity is large enough.

%%%%%%%%%%%%%%%%%%%%%%%%%%%%%%%%%%
\begin{table}[!htbp]
\centering
\caption*{$m=-5$}
\resizebox{\textwidth}{!}
{
        \begin{tabular}{ccccccc}
    \hline\hline
    $n$ &Method&  $\Omega_H=0$  &$\Omega_H=0.5$ & $\Omega_H=1$ & $\Omega_H=5$ & $\Omega_H=10$ \\
    \hline
    $0$ &CFM&     $1.92059 -0.192507i$  & $1.15837 -0.154622i$ & $0.803536 -0.123037i$ & $0.224911 -0.042365i$    & $0.117898 -0.0230009i$\\
        \cline{2-7}
        &WKB&     $1.92059 - 0.192508i$  & $1.15837 - 0.154622i$ & $0.803536 - 0.123037i$ & $0.224911 - 0.0423649i$    & $0.117898 - 0.0230008i$\\
        \hline
    $1$ &CFM&     $1.89952 -0.580296i$  & $1.12353 -0.467438i$ & $0.764354 -0.371602i$ & $0.202946 -0.126386i$    & $0.105041 -0.0683294i$\\
        \cline{2-7}
        &WKB&     $1.89952 - 0.580297i$  & $1.12352 - 0.467302i$ & $0.764257 - 0.371738i$ & $0.202962 - 0.126324i$    & $0.105024 - 0.0682192i$\\
        \hline
    $2$ &CFM&     $1.85899 -0.976231i$  & $1.05457 -0.79145i$ & $0.684919 -0.628419i$ & $0.157842 -0.208351i$    & $0.0785752-0.112125i$\\
        \cline{2-7}
        &WKB&     $1.85898 - 0.976227i$  & $1.0543 - 0.785219i$ & $0.678659 - 0.631877i$ & $0.157978 - 0.206956i$    & $0.0723417-0.11576i$\\
        \hline
    $3$ &CFM&     $1.80226 -1.38504i$  & $0.953952 -1.13541i$ & $0.563948 -0.90167i$ & \textit{non-convergence}    & \textit{non-convergence}\\
        \cline{2-7}
        &WKB&     $1.80214 - 1.38502i$  & $0.958815 - 1.06584i$ & $0.494471 - 0.896412i$ & $0.0867096 - 0.284379i$    & \textit{unreliable}\\
        \hline\hline
\end{tabular}
}
\caption{The first four overtones $n=0,1,2,3$ of QNF at $m=-5$ for different $\Omega_H$. The numerical results are obtained by CFM and WKB. The \textit{non-convergence} in the table means that we can not get convergent numerical results for $n=3$ by CFM at $\Omega_H=5$ and $\Omega_H=10$, and also WKB failed to yield credible results for $n=3$ at $\Omega_H=10$.}\label{tab5}
\end{table}

%%%%%%%%%%%%%%%%%%%%%%%%%%%%%%%
\begin{table}[!htbp]
\centering
\caption*{$m=-10$}
\resizebox{\textwidth}{!}
{
        \begin{tabular}{ccccccc}
    \hline\hline
    $n$ &Method&  $\Omega_H=0$  &$\Omega_H=0.5$ & $\Omega_H=1$ & $\Omega_H=5$ & $\Omega_H=10$ \\
    \hline
    $0$ &AIM&     $3.84704-0.192464i$  & $2.32298-0.154473i$ & $1.61296-0.122871i$ & $0.452634-0.0422895i$    & $0.237420-0.0229599i$\\
        \cline{2-7}
        &WKB&     $3.84704-0.192464i$  & $2.32298-0.154473i$ & $1.61296-0.122871i$ & $0.452634-0.0422895i$    & $0.237419-0.0229592i$\\
        \hline
    $1$ &AIM&     $3.83639-0.578090i$  & $2.30553-0.464301i$ & $1.59347-0.369216i$ & $0.441756-0.126690i$    & $0.231175-0.0687129i$\\
        \cline{2-7}
        &WKB&     $3.83639-0.57809i$   & $2.30553-0.464299i$ & $1.59347-0.369218i$ & $0.441726-0.126693i$    & $0.231032-0.0687109i$\\
        \hline
    $2$ &AIM&     $3.81528-0.965797i$  & $2.27068-0.776801i$ & $1.55434-0.617420i$ & $0.419853-0.210568i$    & $0.221032-0.114104i$\\
        \cline{2-7}
        &WKB&     $3.81528-0.965797i$  & $2.27068-0.776713i$ & $1.55429-0.617518i$ & $0.419159-0.21071i$    & $0.217909-0.114019i$\\
        \hline
    $3$ &AIM&     $3.78412-1.35694i$   & $2.21860-1.09386i$  & $1.49528-0.868883i$ & $0.384825-0.294278i$    & $0.214708-0.164592i$\\
        \cline{2-7}
        &WKB&     $3.78412-1.35694i$   & $2.2185-1.09286i$   & $1.49451-0.869833i$ & $0.381929-0.295194i$    & $0.19664-0.159081i$\\
         \hline\hline
\end{tabular}
}
\caption{The first four overtones $n=0,1,2,3$ of QNF at $m=-10$ for different $\Omega_H$. The numerical results are obtained by AIM and WKB.}\label{tab4}
\end{table}
%%%%%%%%%%%%%%%%%%%%%%%%%%%%%%%

%%%%%%%%%%%%%%%%%%%%%%%%%%%%%%%
%%%%%%%%%%%%%%%%%%%%%%%%%%%%%%%

\begin{figure}[thbp]
\centering
\includegraphics[height=2.4in,width=3.215in]{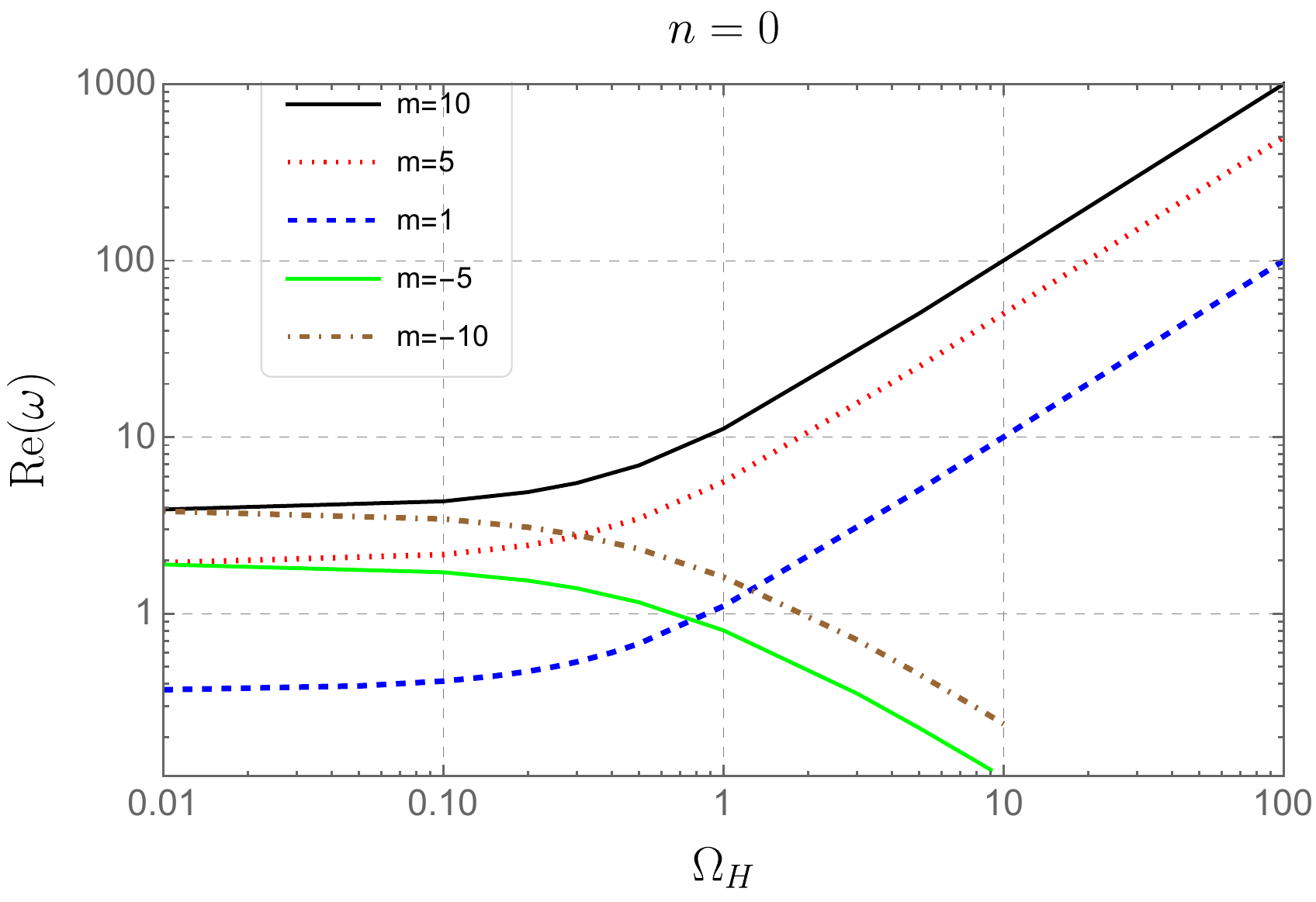}
\includegraphics[height=2.4in,width=3.215in]{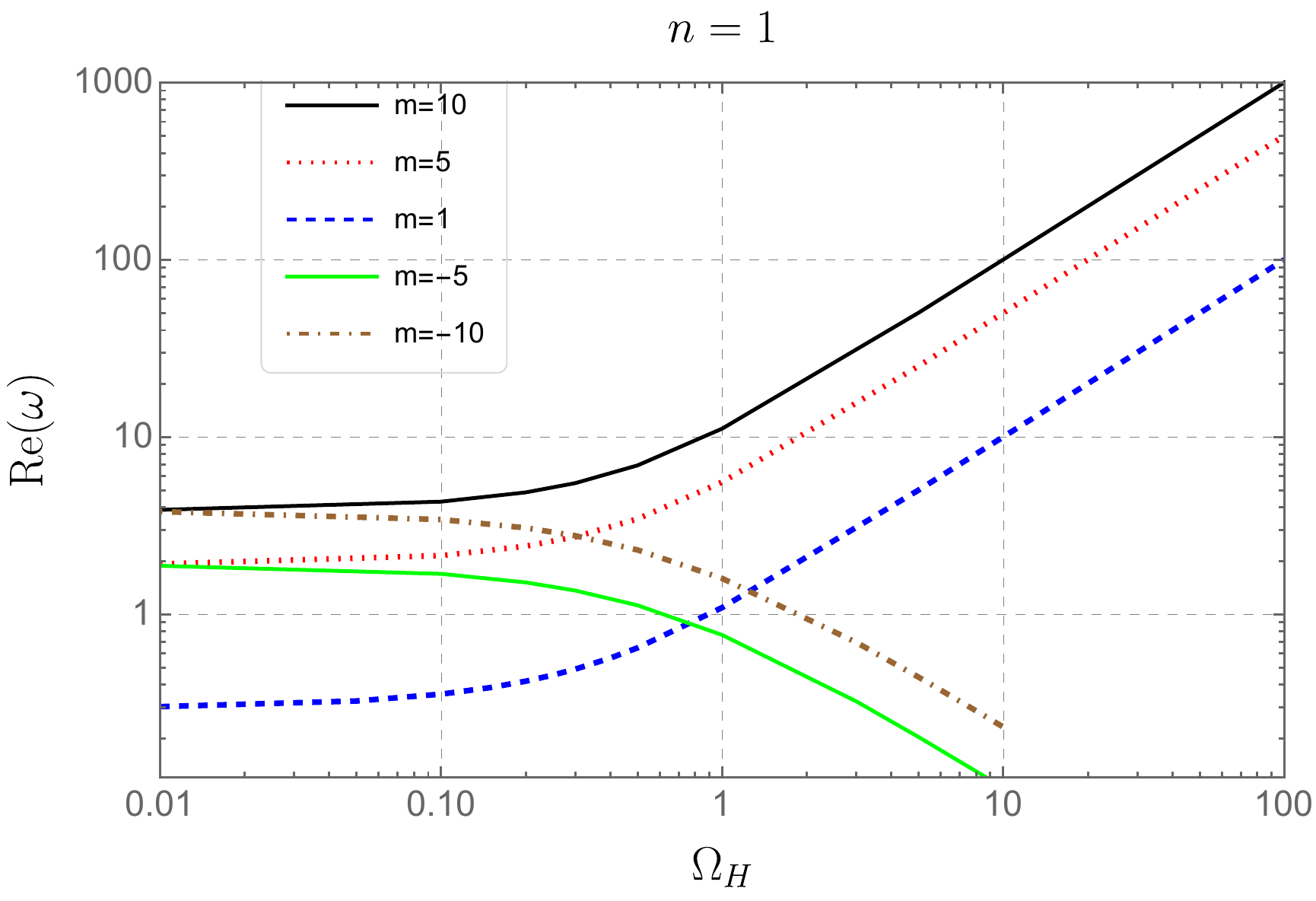}
\includegraphics[height=2.4in,width=3.2in]{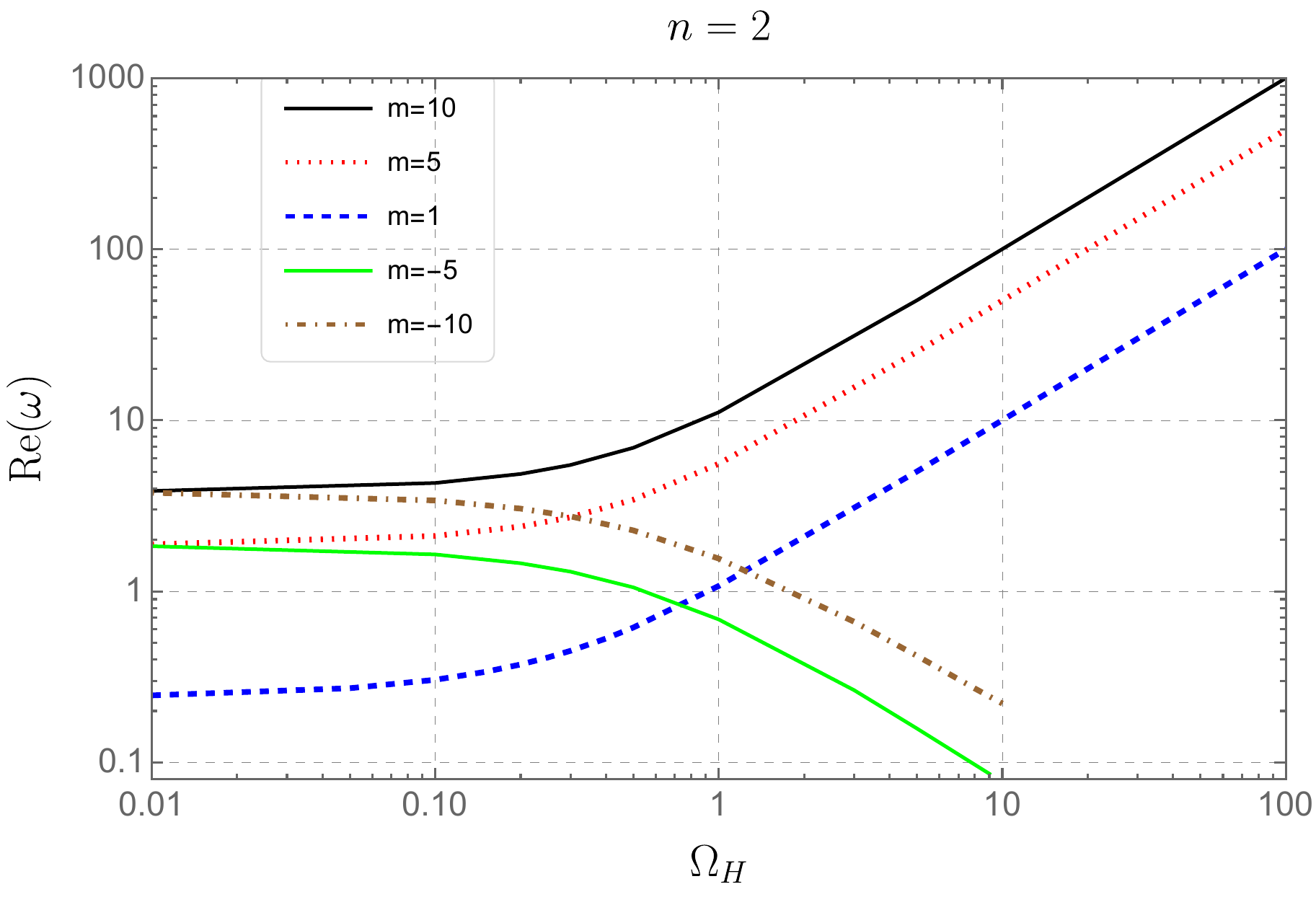}
\includegraphics[height=2.4in,width=3.2in]{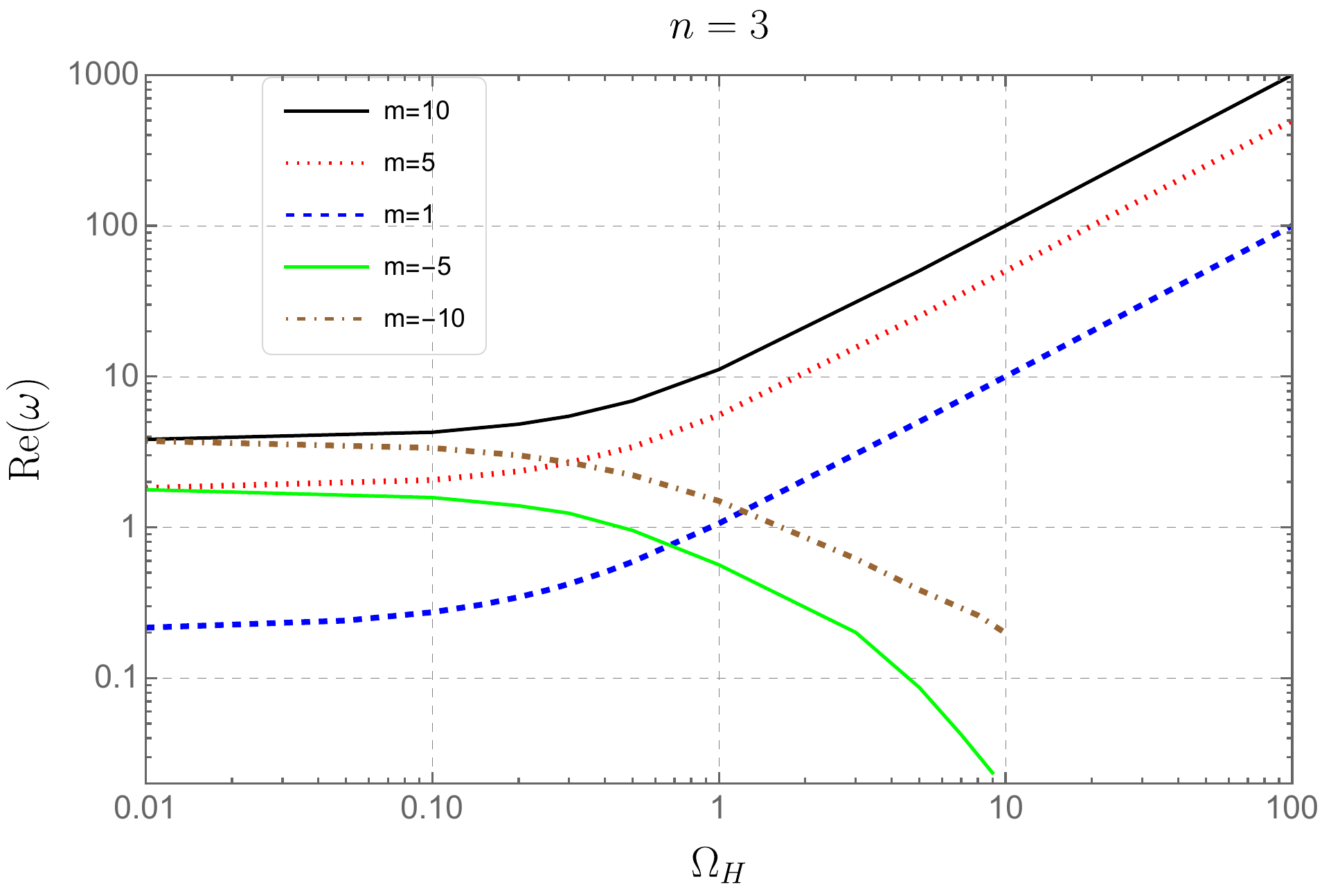}
\caption{The separate comparison of real part $\omega_R$ of overtones from $n=0$ to $3$ between different $m$.\label{fig5}}
\end{figure}

\begin{figure}[thbp]
\includegraphics[height=2.4in,width=3.215in]{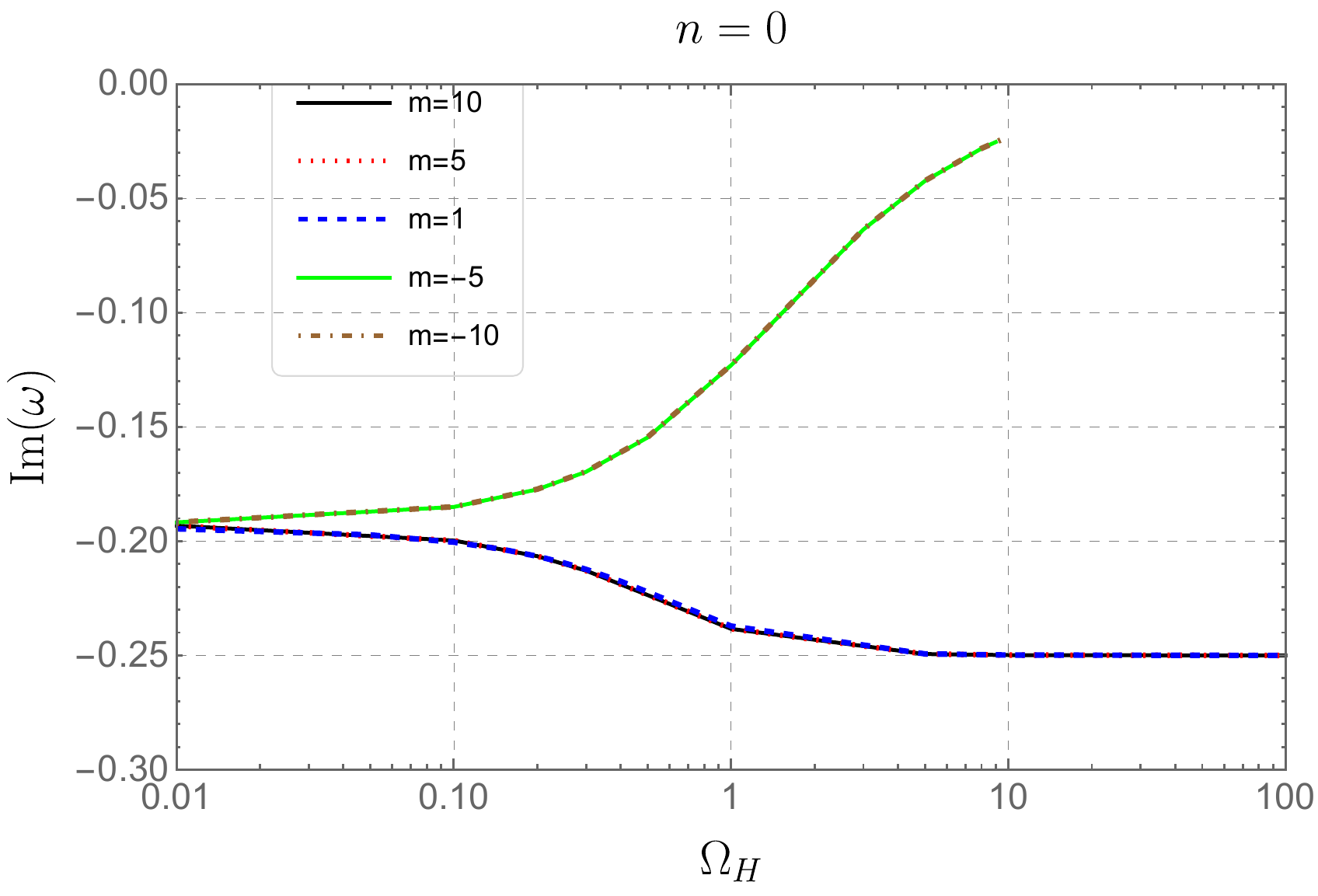}
\includegraphics[height=2.4in,width=3.215in]{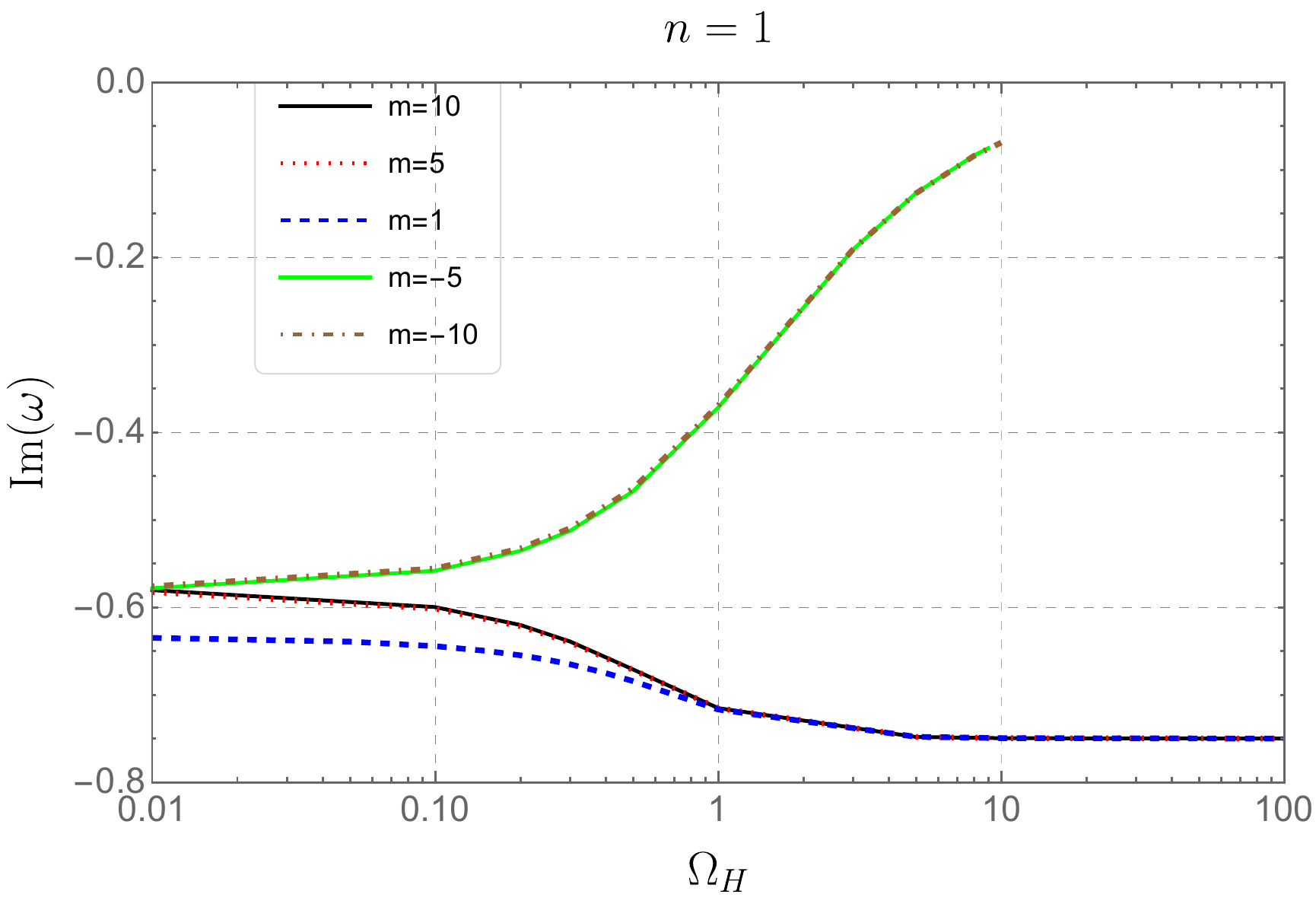}
\includegraphics[height=2.4in,width=3.2in]{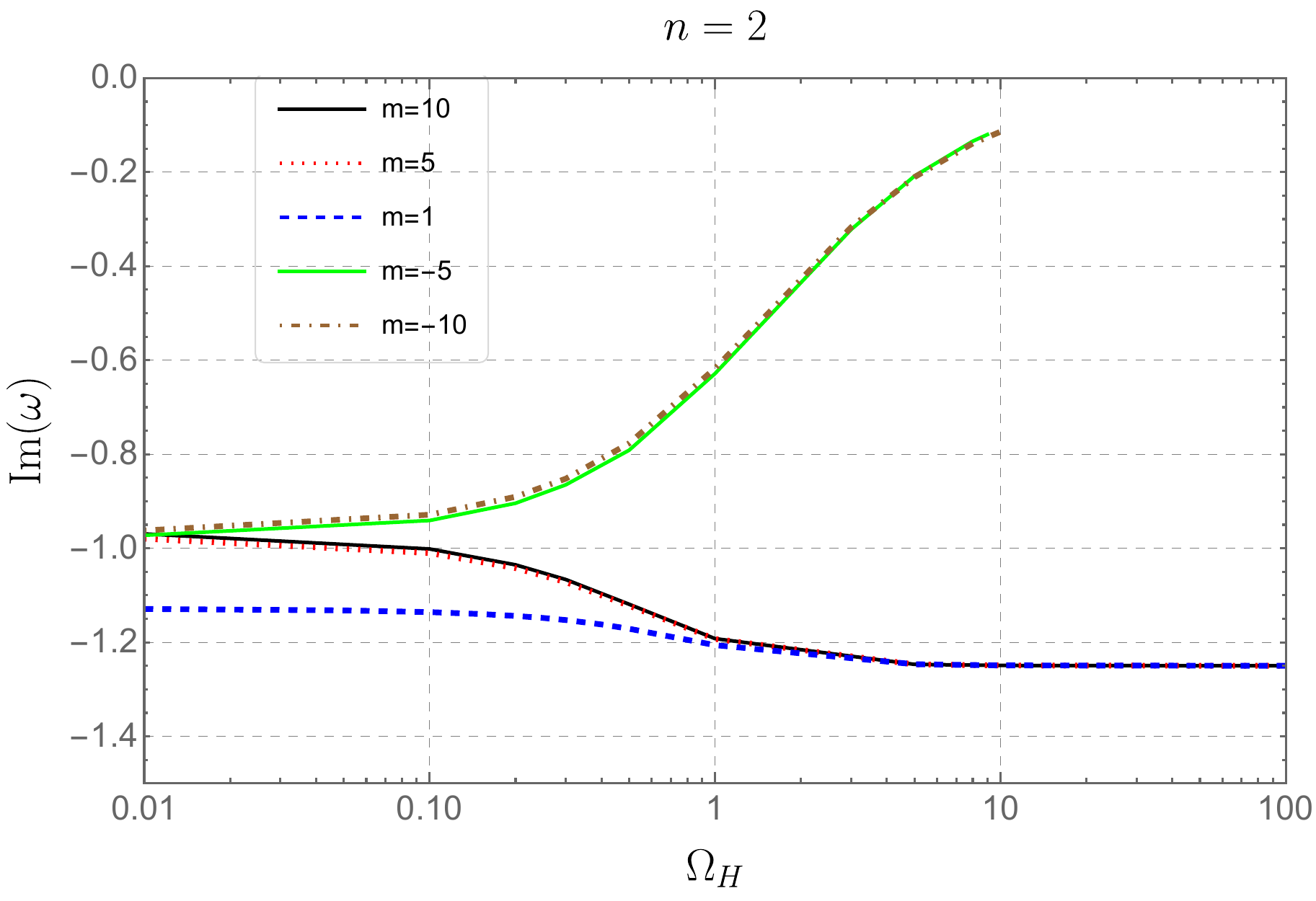}
\includegraphics[height=2.4in,width=3.2in]{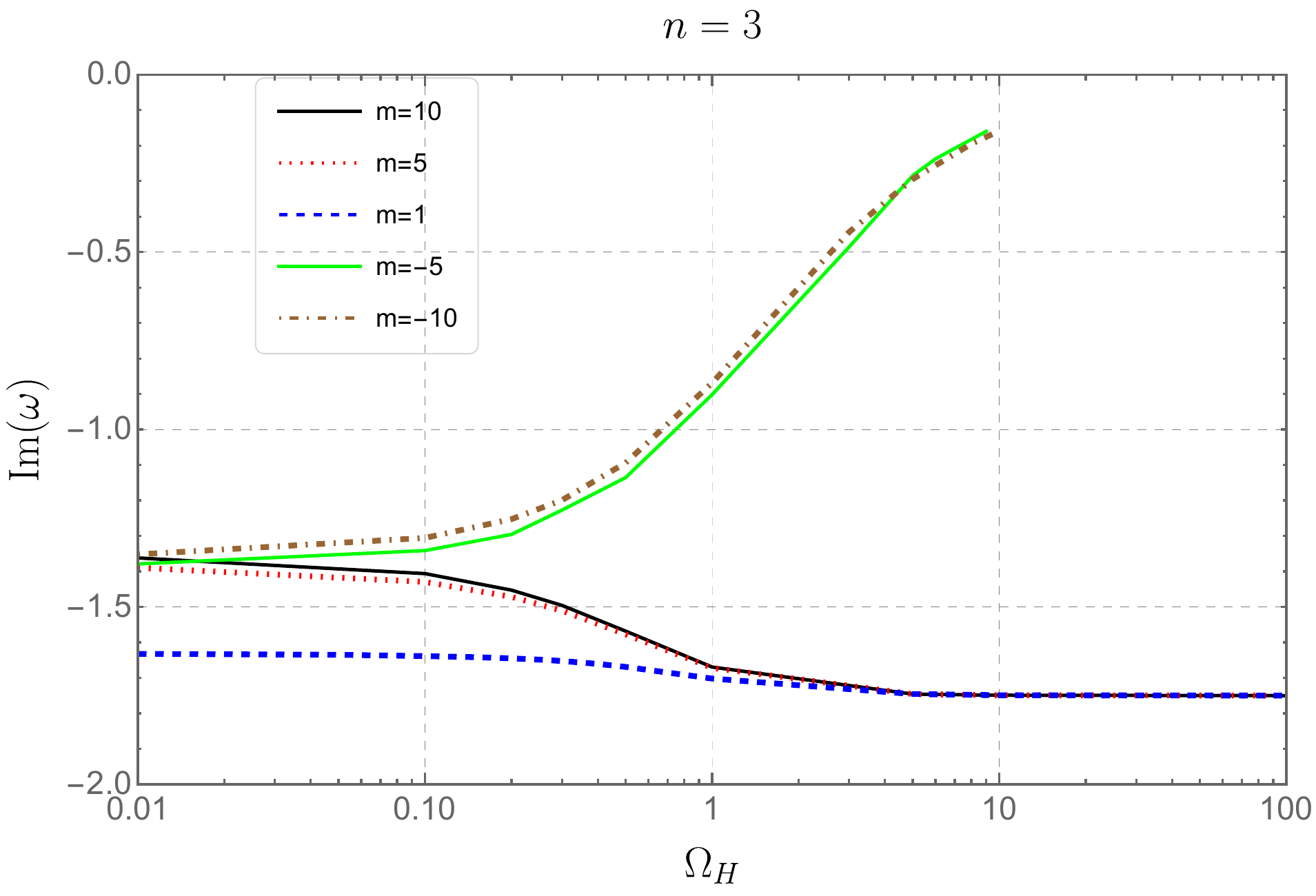}
\caption{The comparison of imaginary part $\omega_I$ of QNF between different $m$, and each plot corresponds to one certain overtone number $n$.\label{fig6}}
\end{figure}

It is natural and necessary to investigate how the QNF is affected by the winding number $m$.
To this end, we separately compare $\omega_R$ and $\omega_I$ between different values of $m$ in Fig.~\ref{fig5} and Fig.~\ref{fig6}, respectively. 
First of all, we observe that the four plots corresponding to four overtones in Fig.~\ref{fig5} appear almost identical, this is attributed to the fact that $\omega_R$ is weakly dependent on overtone number. 
Among the positive values of $m$, it is shown that for any given $\Omega_H$, a larger $m$ is always associated with a higher $\omega_R$, indicating a higher oscillation frequency of QNMs. 
At low values of $\Omega_H$, $\omega_R$ remains almost constant over a small range and then starts to increase slowly. 
After a certain value of $\Omega_H$ (looks like this value is lightly influenced by $m$), the rate of increase in $\omega_R$ becomes much more pronounced, showing a roughly linear relationship on these log-log plots, which implies a power law dependence of $\omega_R$ on $\Omega_H$. 
Note that all the curves are almost parallel to each other, which suggests that the slope is approximately free of $m$. 
As illustrated in Fig.~\ref{fig5}, the slope corresponding to any given overtone number $n$ at high $\Omega_H$ can be determined using linear regression
%we can get the value of slope for any overtone number $n$ at a large $\Omega_H$
%%
\begin{equation}
\frac{\Delta \ln \omega_R}{\Delta\ln \Omega_H}\approx1,
\end{equation}
which directly leads to
\begin{equation}
\ln\omega_R\backsimeq\ln\Omega_H+\ln C_m 
\end{equation}
By taking $\Omega_H=1$, the $m$ dependent constant $C_m$ is determined to be $C_m\approx m$, such that we have
\begin{equation}
 \omega_R\backsimeq m\Omega_H, \quad m>0\, \text{and} \, \Omega_H\to +\infty.
\end{equation}
For a negative winding number, the behavior of QNF diverges significantly from that observed with a positive $m$, as the corresponding $\omega_R$ monotonically decreases when $\Omega_H$ grows and manifests an opposite development to positive winding number case, which indicates a pronounced dependence of $\omega_R$ on both the sign and magnitude of winding number $m$. 
At a specified value of $m$, $\omega_R$ for $-|m|$ keeps to be smaller than that for $|m|$ while they start off at the same value related to $\Omega_H=0$, at which the generalized potential $U(\omega,r)$ depends on $m^2$ such that $\pm m$ will give rise to identical QNF. On the other hand, we can observe that $\omega_R$ for $m=-5$ keeps to be lower than that for $m=-10$ indicating that QNMs with higher $|m|$ always have faster oscillation frequency,  which serves as a common  feature  separately shared by  $\omega_R$ for both negative and positive $m$.

%%%%%%%%%%%%%%%%%%%%%%%%%%%%%%%
%%%%%%%%%%%%%%%%%%%%%%%%%%%%%%%

We now turn our attention to Fig.~\ref{fig6}, which exhibits the comparisons of $\omega_I$ between different $m$. 
One can see that $\omega_I$ for all positive $m$ follows a consistent trend. 
As with the increase of $\Omega_H$, $\omega_I$ for $m>0$ starts to decrease and eventually converges to a certain value determined by the overtone number $n$. 
Naturally, a higher $n$ always gives rise to a greater magnitude of this asymptotic value of $\omega_I$. 
Before the convergence, some other distinctions arising from different $n$ are identified.
For the fundamental modes ($n=0$), all the curves of $\omega_I$ for positive $m$ coincide with each other and this poses a striking contrast with the other plots of higher overtones.
This alignment implies that changes in $m$ within the region of $m>0$ have ignorable impacts on $\omega_I$ for fundamental modes. 
However, for higher overtones, the impacts of positive $m$ show up in the way of inducing deviations of $\omega_I$ among varying $m$ at low values of $\Omega_H$. 
The most significant effects of $m$ is that the curves of $m=1$ is noticeably departed from the curves for $m=5$ and $m=10$. 
For the latter two $m$, a divergence is also formed, just the scale is markedly smaller than previous one. 
Generally, all the deviations in $\omega_I$ arising from $m$ are amplified by increasing $n$. 
This accentuation reveals that among positive $m$, at low values of $\Omega_H$, a higher $m$ will lead to a  negative $\omega_I$ with smaller magnitude, indicating a longer life for QNMs, although this distinction will be eventually eliminated when $\Omega_H$ becomes large.
The behaviors of curves for $m<0$ presents a striking contrast compared to the case of positive $m$, while the same effect of positive $m$ on $\omega_I$ is also observed in this negative $m$ case among which the $\omega_I$ is only slightly affected by $m$ and this kind of effect can  be just mildly amplified for higher overtones. 
When raising $\Omega_H$, it is observed that the curves for $m>0$ have a downward trend, while for $m<0$, the curves have an upward trend. 
Recall that a similar opposite behavior of curves between positive and negative $m$ has also been revealed for $\omega_R$ in Fig.~\ref{fig5}. 
The contrary behaviors of $\omega_I$ between positive and negative $m$, together with the fact that the generalized potential $U(\omega,r)$ is an even function in terms of $m$ at $\Omega_H=0$, such that $\pm m$ has exactly the same QNF and a larger positive $m$ corresponds to a larger $\omega_I$, all these facts finally help us predict that $\omega_I$ for $m=-10$ could be the mode that remains the highest value among all the values of $m$ considered here in the whole  range of $\Omega_H$, as depicted in our figures. 
This result suggests that employing a perturbation optical field with a negative winding number $m$ is preferable for maximizing the lifespan of QNMs. 

%%%%%%%%%%%%%%%%%%%%%%%%%%%%%%%
%%%%%%%%%%%%%%%%%%%%%%%%%%%%%%%

%%%%%%%%%%%%%%%%%%%%%%%%%%%%%%%%%%%%%%%%%%%%%%%%%%%%%%%%%%%%%%%%%%%%%%%%%%%%
\section{Conclusions and Discussions}\label{sec5}

In this paper, we have performed an investigation of the dynamical characteristics of the analog rotating black hole in the photon-fluid model. 
Naturally, these dynamical properties manifest themselves through QNMs, prompting us to calculate the QNF accordingly.
To ensure the reliability of our results, we employed three numerical methods for calculating the QNFs.
Besides the fundamental modes, the overtones are also calculated up to $n=3$ with the intention of acquiring more information of the QNMs. 
We separately discussed the QNF for $m>0$ and $m<0$. 

For $m>0$, we find that the real part $\omega_R$ of QNF for all the overtone number is sensitive to the angular velocity $\Omega_H$ and increases monotonically with the grow of $\Omega_H$, indicating a black hole spinning faster can support QNMs with higher oscillation frequencies. 
Conversely, for the imaginary part $\omega_I$ of QNF, its behavior diverges from that of $\omega_R$. 
With an increase in $\Omega_H$, we find that $\omega_I$ is insensitive to $\Omega_H$ and it monotonically decreases for a while and then eventually approaches a certain negative constant whose value is $n$-dependent. 
Note that a concern arises due to the unlimited angular velocity may cause positive $\omega_I$ implying unstable QNMs, our numerical results suggest this issue is likely unwarranted as $\omega_I$ decreases from a negative value and tends to a negative constant when $\Omega_H$ is sufficiently large.
%this concern has been addressed by our numerical results that $\omega_I$ remains negative for any arbitrarily large $\Omega_H$. 
Additionally, while $\omega_R$ is found to be weakly dependent on the overtone number $n$, $\omega_I$ exhibits a strong correlation on $n$. 
With a higher $n$, $\omega_R$ only mildly drops off and $\omega_I$ experiences a notable reduction. 
When $\Omega_H$ is large, $\omega_R$ displays a approximately linear dependence on $m\Omega_H$, and for neighboring two $\omega_I$ we have $|\Delta\omega_I|\approx0.5$. 
Further more, we find that a larger winding number can increase both $\omega_R$ and $\omega_I$, as $\omega_R\backsimeq m\Omega_H$ for $m>0\, \text{and} \, \Omega_H\to +\infty$. However, $\omega_R$ is notably improved by $m$, whose impacts on $\omega_I$ is slight especially for the fundamental mode. 
An interesting observation is that, the winding number $m$ has noticeable effects on $\omega_R$ but only tiny impacts on $\omega_I$, whereas the overtone number $n$ can only cause mild modifications to $\omega_R$ but notable effects on $\omega_I$. This illustrates that $m$ and $n$ exert opposite influences to QNF.

For $m<0$, strikingly contrasting behaviors of QNF have been observed in comparison to the case of $m>0$.
Specifically, as $\Omega_H$ increases, $\omega_R$ will be monotonically decreased while $\omega_I$ exhibits a continuous increase.
The $\omega_R$ for different overtones decreases with almost the same rate, but for $\omega_I$, a higher overtone demonstrates a notable larger rate of increase, as illustrated in the corresponding semi-log plot.
At high value of $\Omega_H$, all the overtones of $\omega_I$ seem to have trend to converge towards a common certain constant which is close to zero, deviating the relation of $|\Delta\omega_I|\approx0.5$ for $m>0$, but instead converting to $|\Delta\omega_I|\approx0$. 
This trend towards zero for $\omega_I$ suggests the possible existence of arbitrarily long-lived QNMs, also referred to as quasi-resonances.
Considering that when $\Omega_H=0$, the generalized potential $U(\omega,r)$ is an even function in terms of $m$, such that QNF for $\pm |m|$ is expected to be identical at $\Omega_H=0$. 
On the other hand, we have known that when $\Omega_H$ becomes larger the $\omega_R$ will decrease. 
Considering these two observations together, one can easily predict a relation $\omega_{R,-|m|}<\omega_{R,|m|}$ for $m\neq0$, and the same reasoning can also be applied to $\omega_I$ to get $\omega_{I,-|m|}>\omega_{I,|m|}$. 
This simple prediction has been demonstrated in Fig.~\ref{fig5} and Fig.~\ref{fig6}, respectively. 
Consequently, to achieve longer-lived QNMs which brings some convenience for observing, the fluctuating optical field used for generating QNMs should ideally possess a negative winding number $m<0$.

Black hole physics is indisputably important in the modern physics, especially in the attempts of  advancing our understanding of quantum gravity theory. 
Over the past century, black holes have attracted significant attentions, but almost all the researches are limited in theoretical side, due to the extreme hardship of observing the actual black holes in astrophysical environments.
Even if we have detected astrophysical black holes trough gravitational waves, some intriguing effects of black holes, such as Hawking radiation, may still be too faint to be directly detected by current cutting-edge technology of human beings. 
In light of this, simulating analog black holes in laboratory settings seems like a concession to practical limitations, it indeed has proven useful and enlightening in enhancing our comprehension of black hole physics, even including aspects of quantum gravity~\cite{Braunstein:2023jpo}.
Particularly, it is worth noting that the recent exciting advancements in \cite{Svancara:2023yrf} which reported observations of bound states and distinctive analog black hole ringdown signatures in an analog rotating curved spacetime from a giant quantum vortex, indeed brings us a more promising future of studying physics in the black holes spacetime by analogy.

Furthermore, the development of analog gravity has substantially benefited from the extensive application of the Klein-Golden (KG) regime analogy. 
This connection is evident in both theoretical research~\cite{Visser:2005ss} and experimental exploration~\cite{Fedichev:2003id}, underscoring the integral role that the KG regime plays in the emergence and progression of analog gravity models. 
However, on the other hand, deviations from the KG regime are increasingly recognized as crucial for advancing our understanding and experimental validation of analog gravity concepts.
The limitations of the KG equation Eq.~\eqref{eq=kg} are apparent when it comes to the physics beyond Planck scale where the effects of quantum gravity become significant. 
Importantly, phenomena beyond the Planck scale can be effectively explored using microscale models of analog gravity, underscoring the profound significance and motivation behind the study of this field.
In terms of experimental detection of QNMs within the analogous KG framework, established analyses exist~\cite{Torres:2020tzs}. 
Nonetheless, pioneering efforts to move beyond this framework have achieved notable advancements~\cite{Jacquet:2021scv}. 
These endeavors introduce a compelling and challenging direction for future research, inviting further exploration into the rich dynamics of analog gravity models outside the conventional KG regime.

So at the moment it is  worth having a further perspective on the photon-fluid system in the study of analog gravity. Ever since the seminal work of~\cite{PhysRevA.78.063804}, this new analog gravity system has attracted much attentions in community, including research works on acoustic superradiance and superradiant instability \cite{PhysRevA.80.065802,Ciszak:2021xlw}, experimental construction of this analog rotating black hole \cite{Vocke2018}, measurement of superradiance in laboratory \cite{Braidotti:2021nhw}, and the potential applications of the fluids system in the analog simulations of quantum gravity \cite{Braunstein:2023jpo}. In addition, a recent paper \cite{Krauss:2024vst} provided  exciting insights about  employing photon-fluid system to  help resolve
long-standing problems related to quantum gravity, including the black hole information-loss
paradox and the removal of spacetime singularities. All of these advancements have proven that photon-fluid system is a versatile and promising platform which has great potentials to promote our understanding of gravity by analogy.
As for our present work, we analyzed the properties of QNMs for analog black holes in a photon-fluid model, which has been experimentally constructed~\cite{Vocke2018}. 
With these experimental advancements, we are optimistic that the QNMs we have investigated here will soon be observable in apparatus of this photon-fluid model by which provides a novel testbed to black hole physics and theories of gravitation.
As a potential extension, one approach to deviating from the KG regime is to consider the vacuum quantum fluctuations in a photon-fluid model.
According to Ref.~\cite{Jacquet:2021scv,Jacquet:2022vak}, the generation of quantum fluctuations is closely related to the quantum excitations of black hole QNMs. 
At the same time, examining the quantum perturbations of acoustic modes could further our understanding of the quantum effects in microscopic black holes and verify their connections with QNMs.
Additionally, the distinctive characteristics of photon-fluid models offer numerous intriguing questions for future research, as discussed in \cite{Braunstein:2023jpo,Krauss:2024vst}.
For instance, investigating how the breakdown of Lorentz invariance at microscopic scales impacts spacetime could provide valuable insights. 
Quantum Hawking radiation in association with superradiance \cite{Ornigotti:2017yqw} at micro scales are particularly significant, as they underscore the importance of analog gravity research for quantum field theory in curved spacetime.

%%%%%%%%%%%%%%%%%%%%%%%%%%%%%%%%%%%%%%%%%%%%%%%%%%%%%%%%%%%%%%%%%%%%%%%%%%%%
\begin{acknowledgments}
We sincerely thank the anonymous referee whose constructive comments have this work greatly improved. H.L is grateful to  Yanfei for her unwavering and valuable support to his career. This work is supported by the National Natural Science Foundation of China under Grant No.12305071. We also gratefully acknowledge the financial support from Brazilian agencies Funda\c{c}\~ao de Amparo \`a Pesquisa do Estado de S\~ao Paulo (FAPESP), Funda\c{c}\~ao de Amparo \`a Pesquisa do Estado do Rio de Janeiro (FAPERJ), Conselho Nacional de Desenvolvimento Cient\'{\i}fico e Tecnol\'ogico (CNPq), and Coordena\c{c}\~ao de Aperfei\c{c}oamento de Pessoal de N\'ivel Superior (CAPES).
\end{acknowledgments}

%%%%%%%%%%%%%%%%%%%%%%%%%%%%%%%%%%%%%%%%%%%%%%%%%%%%%%%%%%%%%%%%%%%%%%%%%%%%
\appendix
\section{The Connection between Photon-Fluid and Analogue Black Holes}\label{Appendix}

We give a sketch of how to derive the black hole metric in nonlinear optical system. As we have mentioned previously, this derivation has been discussed in~\cite{PhysRevA.78.063804,Ciszak:2021xlw}, while in this Appendix, our purpose is to make a more detailed and concrete discussion on the construction of the analogue rotating black hole spacetime in this optical field system.

To understand how the analogue black hole spacetime is constructed, the starting point is to see what we can get from the hydromechanics. An insightful discovery by Unruh~\cite{Unruh:1980cg,Unruh:1994je} is that an analogue black hole spacetime can be built from the following fundamental equations
\vspace{-20pt}
\begin{subequations}\label{ap_eq13}
\begin{gather}
\partial_{t} \rho +\nabla\cdot(\rho\,\mathbf{v})=0, \quad\quad \text{Equation of continuity,}\\ \label{ap_eq2}
\rho\frac{d\mathbf{v}}{dt}\equiv\rho\left[\partial_{t}\mathbf{v}+(\mathbf{v}\cdot\nabla)\mathbf{v}\right]=\mathbf{f}, \quad\quad \text{ Euler's equation,}
\end{gather}
\end{subequations}
which describe the dynamics of fluids.  
In this equation, $\rho$ is the density of the fluids and we have used boldface to denote the fluids velocity vector $\mathbf{v}$ and the force vector $\mathbf{f}$ acted on the fluids. 
So the Euler equation is just equivalent to Newton's second law $\mathbf{f}=m\,\mathbf{a}$ with $\mathbf{f}$ being the force applied to a small lumps of the fluid. 

We now focus on the Euler's equation. From the knowledge of the vector analysis, we have identity
\begin{equation}
\mathbf{v}\times(\nabla\times\mathbf{v})=\frac{1}{2}\nabla v^2-(\mathbf{v}\cdot\nabla)\mathbf{v},
\end{equation}
where $v^2=\mathbf{v}\cdot\mathbf{v}$. 
By assuming the fluid to be inviscid (zero viscosity) which means that the only forces present being those due to pressure $p$. 
Then for the force density, we have
\begin{equation}
\mathbf{f}=-\nabla p.
\end{equation}
With these conditions, we have the following Euler's equation
\begin{equation}\label{ap_eq1}
\partial_{t}\mathbf{v}=\mathbf{v}\times(\nabla\times\mathbf{v})-\frac{1}{2}\nabla v^2-\frac{\nabla p}{\rho}.
\end{equation}
Furthermore, we assume that the fluids is locally irrotational which directly leads to $\nabla\times\mathbf{v}=0$ (vorticity free), and it implies that the velocity vector is the gradient of some scalar $\phi$, i.e. $\mathbf{v}=-\nabla \phi$. 
On the other hand, we also require the fluid to be barotropic which means that the density $\rho$ is a function of pressure $p$ only, such that it is possible to define a function $h(p)$ as
\begin{equation}
h(p)=\int_{0}^{p}\frac{dp'}{\rho(p')}=\int_{x(0)}^{x(p)}\frac{\nabla p'\cdot d\mathbf{x}}{\rho(p')}=\int_{x(0)}^{x(p)} \nabla h(p')\cdot d{\mathbf{x}},
\end{equation}
which shows that
\begin{equation}\label{ap_eq4}
\nabla h=\frac{\nabla p}{\rho}.
\end{equation}
For more details, one can refer to \cite{Barcelo:2005fc} to have a comprehensive understanding on analogue gravity and the prodedures of the derivation of black hole metric from Eq.~\eqref{ap_eq13}.

The starting point of finding a black hole metric in optical system is from the equation which describes the slowly varying envelope of the optical field. 
The dynamics of the photon-fluid in our consideration is governed by the nonlinear Schrödinger equation (NSE)
\begin{equation}
\frac{\partial E}{\partial z}=\frac{i}{2k}\nabla^2 E-i\frac{k n_2}{n_0}E |E|^2,\label{ap_eq12}
\end{equation}
where $z$ is the propagation direction, $|E|^2$ is the optical field intensity and $\nabla^2 E$ is defined with respect to the transverse coordinates $(x, y)$, $i$ is the imaginary unit, $k=2 \pi n_0/\lambda$ is the wave number, $\lambda$ is the laser wavelength in vacuum, $n_0$ is the linear refractive index, $n_2>0$ is the material nonlinear coefficient which characterizes the intensity dependent refractive index $n=n_0-n_2|E|^2$ of the self-defocusing media. 
As a consequence, light propagating in a self-defocusing medium induces a local negative bending of the refractive index which, in turn, affects the light beam itself. 
At a microscopic level this can be described in terms of an atom-mediated repulsive interaction between photons which leads to the formation of a “photon fluid”, such that we get a flavor of fluids in optical system here.

We now start to reveal the hidden links between Eq.~\eqref{ap_eq12} describing optical field and Eq.~\eqref{ap_eq13} describing fluids dynamics. 
The first step in this task is to apply the Madelung transformation to write the  complex EM field $E$ in terms of its amplitude and phase
\begin{equation}
E=\rho^{\frac{1}{2}}e^{i\phi},
\end{equation}
and by this formula of $E$, we would like to investigate Eq.~\eqref{ap_eq12} term by term. 
The term on the left hand side is expressed as
\begin{equation}
\partial_z E=\partial_z(\rho^{\frac{1}{2}}e^{i\phi})=i \rho^{\frac{1}{2}}e^{i\phi}\partial_z\phi+\frac{1}{2}\rho^{-\frac{1}{2}}e^{i \phi}\partial_z\rho.
\end{equation}
The first term on the righthand side is
\begin{equation}
\begin{split}
\nabla^2 E&=\nabla^2 (\rho^{\frac{1}{2}}e^{i\phi})=\rho^{\frac{1}{2}}\nabla^2e^{i\phi}+2\nabla\rho^{\frac{1}{2}}\cdot\nabla e^{i\phi}
+e^{i\phi}\nabla^2\rho^{\frac{1}{2}}\\
&=\rho^{\frac{1}{2}}e^{i\phi}(i\nabla^2\phi-(\nabla\phi)^2)+i\rho^{-\frac{1}{2}}e^{i\phi}\nabla
\rho\cdot\nabla\phi+e^{i\phi}\nabla^2\rho^{\frac{1}{2}}\\
&=\frac{1}{\alpha}ie^{i\phi}\rho^{-\frac{1}{2}}[\nabla\rho\cdot\nabla (\alpha\phi)+\rho\nabla^2(\alpha\phi)]-\frac{1}{\alpha^2}\rho^{\frac{1}{2}}e^{i\phi}(\nabla(\alpha\phi))^2+e^{i\phi}\nabla^2\rho^{\frac{1}{2}}\\
&=\frac{1}{\alpha}ie^{i\phi}\rho^{-\frac{1}{2}}\nabla\cdot(\rho\mathbf{v})-\frac{1}{\alpha^2}\rho^{\frac{1}{2}}e^{i\phi}\mathbf{v}^2+e^{i\phi}\nabla^2\rho^{\frac{1}{2}},
\end{split}
\end{equation}
where we have defined ``fluids'' velocity  $\mathbf{v}=\alpha\nabla\phi\equiv\nabla\psi$ and $\alpha=\frac{c}{kn_0}$ (here $c$ is the speed of the light). 
The last term on righthand side is
\begin{equation}
E|E|^2=\rho^{\frac{3}{2}}e^{i\phi}.
\end{equation}
We substitute these obtained equations in Eq.~\eqref{ap_eq12} and we get a pair of equations
\begin{align}
&\frac{1}{2}\rho^{-\frac{1}{2}}e^{i\phi}\partial_z\rho=-\frac{1}{2\alpha k}\rho^{-\frac{1}{2}}e^{i\phi}\nabla\cdot(\rho\mathbf{v})\label{ap_eq14}\\
&i\rho^{\frac{1}{2}}e^{i\phi}\partial_z\phi=\frac{i}{2k}\left[-\frac{1}{\alpha^2}\rho^{\frac{1}{2}}e^{i\phi}\mathbf{v}^2+e^{i\phi}\nabla^2\rho^{\frac{1}{2}}\right]
-i\frac{kn_2}{n_0}\rho^{\frac{3}{2}}e^{i\phi},\label{ap_eq15}
\end{align}
The Eq.~\eqref{ap_eq14} leads to
\begin{equation}
k\alpha\frac{\partial \rho}{\partial z}+\nabla\cdot(\rho\mathbf{v})=\frac{c}{n_0}\frac{\partial \rho}{\partial z}+\nabla\cdot(\rho\mathbf{v})=0.
\end{equation}
Eq.~\eqref{ap_eq15} leads to
\begin{equation}
\frac{c}{n_0}\frac{\partial(\alpha\phi)}{\partial z}+\frac{1}{2}\mathbf{v}^2-\frac{\alpha^2}{2}\frac{\nabla^2\rho^{\frac{1}{2}}}{\rho^{\frac{1}{2}}}+\frac{k^2\alpha^2 n_2}{n_0}\rho=0.
\end{equation}
We define $t=\frac{n_0}{c}z$ and note that $\psi=\alpha \phi$, which results in
\begin{align}
&\frac{\partial \rho}{\partial t}+\nabla\cdot(\rho\mathbf{v})=0,\label{ap_eq16}\\
&\frac{\partial \psi}{\partial t}+\frac{1}{2}\mathbf{v}^2+\frac{c^2n_2}{n_0^3}\rho-\frac{1}{2}\frac{c^2}{k^2n_0^2}\frac{\nabla^2\rho^{\frac{1}{2}}}{\rho^{\frac{1}{2}}}=0.\label{ap_eq17}
\end{align}
Eq.~\eqref{ap_eq16} is just the equation of continuity which is in exact the same form as that in fluids dynamics, and Eq.~\eqref{ap_eq17} is the Euler's equation. 
The last term in Eq.~\eqref{ap_eq17} represents quantum pressure which has no analogy in classical fluid dynamics. 
In optics, it is a direct consequence of the wave nature of light (it arises from the diffraction term $\nabla^2E$) and is significant in rapidly varying and/or low-intensity regions such as dark-soliton cores and close to boundaries.

When the quantum pressure is negligible, then Eq.~\eqref{ap_eq17} has the exact same form of the Euler's equation for fluid dynamics. 
By comparison, it is clear to see that $h(p(\rho))\equiv\frac{c^2n_2}{n_0^3}\rho$, and take Eq.~\eqref{ap_eq4} into consideration, we can get the relation between $p$ and $\rho$,
\begin{equation}
P=\frac{c^2n_2\rho^2}{2n_0^3},
\end{equation}
such that the speed of sound $c_s$ can be obtained as 
\begin{equation}
c_s^2=\frac{\partial P}{\partial \rho}\bigg|_{\rho=\rho_0}=\frac{c^2n_2\rho_0}{n_0^3}.
\end{equation}
As we have obtained the same equation of continuity and Euler's equation in optical system as that in hydromechanics, the connections between the photon-fluid and the analog gravity is clear now, since these two fundamental equations are the starting point of constructing analog gravity in fluids.

%%%%%%%%%%%%%%%%%%%%%%%%%%%%%%%%%%%%%%%%%%%%%%%%%%%%%%%%%%%%%%%%%%%%%%%%%%%%
\bibliographystyle{JHEP}
%\bibliographystyle{apsrev4-1}
%注意tex文件名不能有空格否则参考文献识别不出来！！！！！！！！！！！！！！！！
\bibliography{References_Analog_BH}

\end{document}